\documentclass[aps,prd,onecolumn,superscriptaddress,nofootinbib]{revtex4-2}

\usepackage{graphicx}
\usepackage{amsmath,amssymb}
\usepackage{bm}
\usepackage{xcolor}
\usepackage{hyperref}
\usepackage{orcidlink}
\usepackage{booktabs}
\usepackage{multirow}
\usepackage{siunitx}
\usepackage{tikz}
\usetikzlibrary{arrows.meta, positioning, calc}

\hypersetup{colorlinks=true,linkcolor=blue,citecolor=blue,urlcolor=blue}

\begin{document}

\title{Probing Nucleon Spin Structure with a Polarized Gamma Beam from Compton Backscattering at FCC-ee}

\author{A.~C.~Canbay\,\orcidlink{0000-0003-4602-473X}}
\email{Contact author: acanbay@ankara.edu.tr}
\affiliation{Ankara University, Ankara, Türkiye}

\author{S.~Sultansoy\,\orcidlink{0000-0003-2340-748X}}
\affiliation{TOBB University of Economics and Technology, Ankara, Türkiye}

\author{F.~Zimmermann\,\orcidlink{0000-0001-9787-8917}}
\affiliation{European Organization for Nuclear Research (CERN), Geneva, Switzerland}

\date{\today}

\begin{abstract}
We present a kinematic and optical design of a high-energy polarized gamma-ray facility based on Compton backscattering of lasers against the FCC-ee electron beams in its $Z$, $WW$, $ZH$ and $t\bar{t}$ modes. The conversion point is located in the FCC-ee full-energy booster, allowing parasitic CBS operation without dedicated interaction-point optics. Saturating the safe value of the kinematic parameter $\kappa = 4.35$ in each mode fixes the laser wavelength and yields backscattered photons up to $\omega_{\max} = 148$~GeV.  The conversion point operates on the booster cycle structure in two modes: fully parasitically on the 0.1~s extraction flat-tops of the top-up ramps ($f_{\rm CBS}=10^{-8}$ per bunch crossing, cumulative electron loss $3\times10^{-6}$ per cycle), and, as the baseline, in dedicated extended-flat-top fills inside the idle windows between top-up cycles ($f_{\rm CBS}=10^{-6}$, beam loss below 1\% per cycle, no impact on the collider luminosity); the corresponding operational laser pulse energies are in the sub-millijoule to few-millijoule range. Polarized photon selection is performed event-by-event via a pair spectrometer that reconstructs $E_\gamma$ on the high-energy Compton edge; for the unpolarized booster beam the Compton polarization transfer limits the achievable band-averaged circular polarization, and the selection is set to $|\langle S_{2}\rangle| = 0.90$. We project the resulting sensitivity to the polarized gluon distribution $\Delta g(x)$ through open-charm photoproduction $\gamma p \to c\bar{c}X$ on an NH$_{3}$ dynamic-nuclear-polarization target, including next-to-leading-order QCD corrections via $K$-factors and propagating polarized-PDF uncertainties through the $100$ Monte Carlo replicas of NNPDFpol2.0. The projected total precision on $\Delta g(x)/g(x)$ is $\delta(\Delta g/g)_{\rm tot}\simeq 1.8$--$3.0\times 10^{-2}$, a factor of $\sim 4$--$7$ smaller than the total uncertainty of the most precise existing direct world measurement (HERMES, dominated by Monte-Carlo model uncertainties), with four distinct values of $\langle x\rangle$ in the medium-$x$ region $0.07\leq x\leq 0.19$. The proposed facility would set the dominant constraint on the polarized gluon distribution in the medium-$x$ region, complementary to the low-$x$ reach of the Electron--Ion Collider.
\end{abstract}


\maketitle

\section{Introduction}
The spin structure of the proton remains one of the fundamental open problems of quantum chromodynamics (QCD). In the naive quark-parton model, one expects the total quark helicity to account for the full proton spin, yielding $\Delta\Sigma = 1$. This expectation was overturned in 1988 when the European Muon Collaboration (EMC) measured the polarized proton structure function $g_{1}^{p}$ in deep-inelastic scattering (DIS) and found $\Delta\Sigma$ to be far smaller than predicted~\cite{Ellis1974,Ashman1988}. Subsequent measurements by the Spin Muon Collaboration (SMC)~\cite{Adeva1993,Adams1994} and the SLAC E-142 and E-143 experiments~\cite{Anthony1993,Abe1995} confirmed this finding; modern global analyses converge on $\Delta\Sigma\approx 0.20$--$0.30$ at $Q^{2}=4$~GeV$^{2}$~\cite{Aidala2013}, in direct violation of the Ellis--Jaffe sum rule~\cite{Ellis1974}, which has since become known as the ``proton spin crisis''.

Within perturbative QCD, it was shown independently by Altarelli and Ross~\cite{Altarelli1988} and by Efremov and Teryaev~\cite{Efremov1990} that the axial anomaly allows the gluon spin to contribute to the measured quark helicity sum, thereby modifying the Ellis--Jaffe prediction. The proton spin budget is decomposed as
\begin{equation}
\frac{1}{2} = \frac{1}{2}\Delta\Sigma + \Delta G + L_{q,g},
\label{eq:spin_decomp}
\end{equation}
where $\Delta G$ is the first moment of the polarized gluon distribution and $L_{q,g}$ denotes the sum of orbital angular momenta of all partons. Despite this decomposition being well established, the gluon contribution $\Delta g(x)$ remains poorly constrained, with theoretical estimates spanning a wide range~\cite{Gluck1990,GRSV2001,Aidala2013}, particularly in the moderate-$x$ region. Proposals based on polarized proton--proton collisions~\cite{Ramsey1988,Berger1989}, polarized photoproduction with large-$p_{T}$ hadron pairs~\cite{Peralta1994}, charm photoproduction~\cite{Keller1994}, or semi-inclusive muon-proton charm production~\cite{Watson1982} access only products of distribution functions rather than $\Delta g(x)$ in isolation, preventing a direct and unambiguous measurement.

Scattering high-energy polarized photon beams off polarized nuclear targets circumvents this limitation. The Compton backscattering (CBS) technique~\cite{Ginzburg1983,Telnov1990} converts circularly polarized laser photons ($\omega_{0}\sim$~eV) into energetic, nearly fully polarized gamma-ray beams through inverse Compton scattering off ultra-relativistic electrons. The kinematic parameter $\kappa = 4 E_{e}\omega_{0}/m_{e}^{2}$, where $E_{e}$ is the electron beam energy and $m_{e}$ the electron mass, governs the energy boosting; the maximum scattered photon energy is $\omega_{\max} = \kappa E_{e}/(\kappa+1)$. This approach was first proposed for nucleon spin studies by Alekhin, Borodulin, and Sultanov~\cite{Alekhin1993}. Atag \emph{et al.} extended this proposal to ring-type accelerators including LEP, TRISTAN, and HERA~\cite{Atag1995,Atag1996}, deriving spin asymmetries for $J/\psi$ photoproduction and two-hadron production and projecting an annual integrated luminosity of $\sim 1$~fb$^{-1}$. Alekhin \emph{et al.} subsequently extended the program to the TESLA linear collider~\cite{Alekhin1999}, demonstrating that open-charm photoproduction ($D$-meson tagging) at $250$~GeV provides direct access to $\Delta g/g$ through the photon-gluon fusion channel. We note that all of these proposals remained projections; the operating CBS gamma sources realised to date, such as ROKK-1M at VEPP-4M~\cite{ROKK1M}, HI$\gamma$S at Duke~\cite{HIGS}, and ELI-NP at IFIN-HH~\cite{ELINP}, reach photon energies of at most $\sim 1.5$~GeV, two orders of magnitude below the open-charm photoproduction threshold and therefore well outside the kinematic range considered in the present work.

\begin{figure}[!htbp]
\centering
\includegraphics[width=0.80\linewidth]{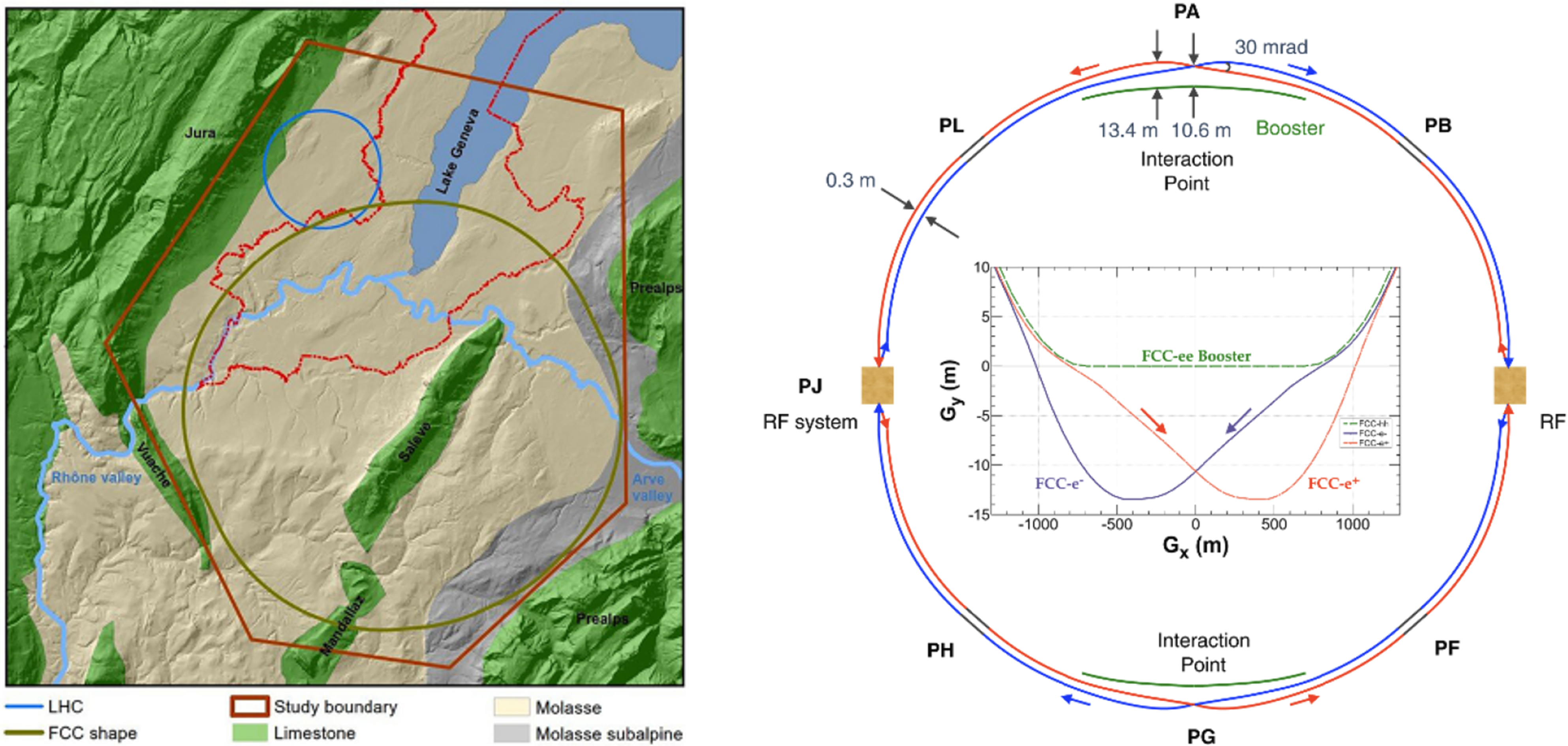}
\caption{Layout of the FCC, with a circumference of 90.7~km, showing the four collision points and the four technical insertions (left; reproduced from the FCC Feasibility Study Report, Vol.~3~\cite{FCCFSRvol3}). Arrangement of the FCC-ee collider rings and the full-energy booster along the FCC-hh footprint, with a zoomed view of the beam trajectories across an interaction point (right; from the FCC-ee Conceptual Design Report~\cite{FCC2019}, Fig.~2.1).}
\label{fig:fcc_layout}
\end{figure}

The Future Circular Collider in its electron--positron phase, FCC-ee, is designed to operate near CERN in a 91~km circumference tunnel, covering center-of-mass energies from approximately 91 to 365~GeV across four main operating modes ($Z$, $WW$, $ZH$, $t\bar{t}$)~\cite{FCC2019,FCC2025,FCCFSRvol2,FCCFSRvol3}. Its layout is shown in Fig.~\ref{fig:fcc_layout} and its key design parameters are summarized in Table~\ref{tab:fccee_params}. The high beam current, ultralow emittance, and large number of bunches make FCC-ee an unparalleled platform for a CBS-based polarized gamma-ray facility, provided the Compton interaction is operated parasitically so as to preserve the nominal collider luminosity. Most recently, Agapov \emph{et al.}~\cite{Agapov2026} discussed potential CBS-based photon source applications at FCC-ee as part of a broad survey of auxiliary science opportunities, identifying it as a promising avenue but stopping short of any quantitative photon-beam design, Monte Carlo characterization, or spin-physics projection. The present work addresses this gap by providing the first complete kinematic and optical design across all four FCC-ee operating modes, validated by full QED simulations, with explicit treatment of (i) the Compton fraction compatible with parasitic operation of the collider, (ii) the role of the downstream collimator system as a radiation envelope and shower-cleaning device (removal of the $e^{+}e^{-}$ showers generated by the off-axis photon halo in the primary-collimator jaws) rather than a polarization filter, (iii) the event-level photon-energy reconstruction by the pair spectrometer that selects the polarized Compton-edge band, and (iv) the forward beam hole and beam dump required to absorb the non-interacting photon flux.

\begin{table*}[t]
\centering
\caption{Key design parameters of FCC-ee for each operating mode~\cite{FCC2025,FCCFSRvol2}. Beam parameters at the CBS conversion point in the booster are given separately in Table~\ref{tab:booster_params}.}
\label{tab:fccee_params}
\begin{tabular}{lcccc}
\toprule
Parameter & $Z$ & $WW$ & $ZH$ & $t\bar{t}$ \\
\midrule
Center-of-mass energy $\sqrt{s}$ [GeV]              & 91.2 & 160  & 240  & 365   \\
Beam energy $E_e$ [GeV]                              & 45.6 & 80   & 120  & 182.5 \\
Particles per bunch $N_e$ ($\times 10^{11}$)         & 2.18 & 1.38 & 1.69 & 1.58  \\
Number of bunches $n_b$                              & 11200 & 1780 & 440 & 60   \\
Beam current [mA]                                    & 1283 & 135  & 26.8 & 5.1  \\
Luminosity/IP ($10^{34}$~cm$^{-2}$s$^{-1}$)         & 145  & 20   & 7.5  & 1.41 \\
RMS bunch length (SR) [mm]                           & 5.53 & 3.46 & 3.26 & 1.91 \\
\bottomrule
\end{tabular}
\end{table*}

In this paper, we present a complete kinematic and optical design of a CBS-based high-energy polarized gamma-ray facility driven by FCC-ee electron beams. Monte Carlo simulations performed with the CAIN code~\cite{CAIN1995} are used to characterize the photon beam flux and polarization profile across all FCC-ee operating modes. The interaction of the polarized gamma beam with a polarized fixed target and the resulting spin-asymmetry projections are then evaluated through an analytic perturbative-QCD treatment of open charm photoproduction, using polarized and unpolarized parton distribution functions, following the standard methodology established by Bojak and Stratmann~\cite{BojakStratmann1999} and the COMPASS open-charm program~\cite{COMPASS2013}. The paper is organized as follows. Section~\ref{sec:setup} describes the experimental setup and the overall layout of the facility, including the parasitic CBS operating point, the collimator-and-dipole shower-cleaning system, the pair-spectrometer-based event-level polarization selection, and the downstream beam dump. Section~\ref{sec:kinematics} determines the optimal laser parameters for each operating mode from the kinematic constraints on $\kappa$ and presents the CBS conversion point design. Section~\ref{sec:cbs_simulation} presents the CAIN Monte Carlo simulations and the resulting collimator geometry and operational photon flux. Section~\ref{sec:asymmetry} discusses the open-charm photoproduction analysis and the spin-asymmetry projections, and translates them into a projected determination of $\Delta g(x)/g(x)$ compared against the existing world data.

\section{Experimental Setup}
\label{sec:setup}

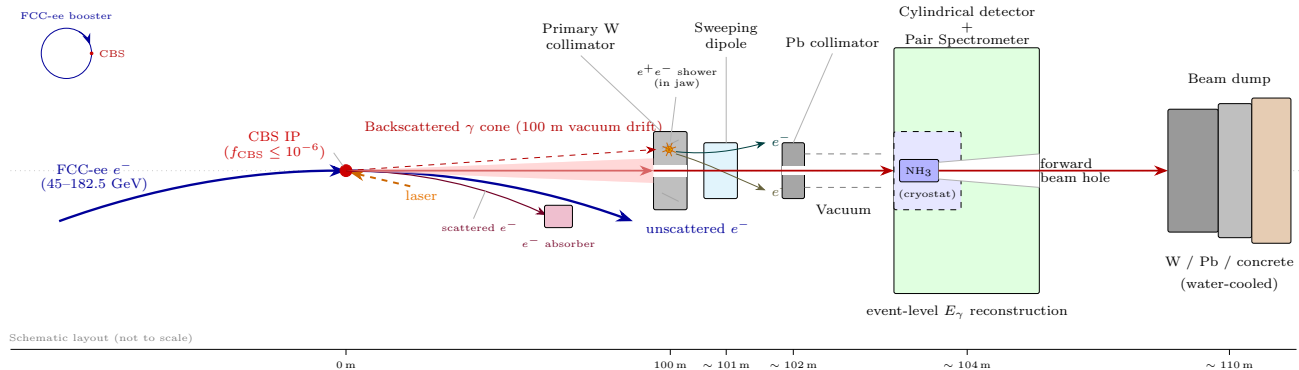
\begin{figure*}[!htbp]
\centering
\resizebox{\textwidth}{!}{


\begin{tikzpicture}[
    x=1.0cm, y=1.0cm,
    >=Stealth,
    every node/.style={font=\small},
    arrow_beam/.style={->, line width=1.2pt, blue!60!black},
    arrow_gamma/.style={->, line width=0.9pt, red!70!black},
    arrow_laser/.style={->, line width=1.0pt, orange!80!black, dashed},
    arrow_eplus/.style={->, line width=0.4pt, blue!50!black},
    arrow_eminus/.style={->, line width=0.4pt, red!50!black},
    component_fill/.style={draw, line width=0.4pt, rounded corners=1pt},
]

\draw[gray!50, dotted, line width=0.4pt] (-6, 0) -- (18, 0);

\begin{scope}[shift={(-5, 2.1)}, scale=0.45]
  \draw[blue!60!black, line width=0.5pt] (0, 0) circle (1.0);
  \draw[arrow_beam, line width=0.6pt]
    ([shift=(60:1.0)]0,0) arc[start angle=60, end angle=20, radius=1.0];
  \node[blue!60!black, font=\tiny, anchor=south]
    at (0, 1.1) {FCC-ee booster};
  \filldraw[red!80!black] (0:1.0) circle (1.5pt);
  \node[red!70!black, font=\tiny, anchor=west]
    at (0:1.05) {CBS};
\end{scope}

\draw[arrow_beam]
  ([shift=(110:15.0)]0,-15.0) arc[start angle=110, end angle=90, radius=15.0];
\draw[arrow_beam]
  ([shift=(90:15.0)]0,-15.0) arc[start angle=90, end angle=70, radius=15.0];

\node[blue!60!black, font=\scriptsize, align=center] at (-4.5, 0.05)
  {FCC-ee $e^-$};
\node[blue!60!black, font=\scriptsize, align=center] at (-4.5, -0.25)
  {(45--182.5 GeV)};

\node[blue!60!black, font=\scriptsize, anchor=west] at (5.25, -1.0)
  {unscattered $e^-$};

\draw[arrow_laser]
  (1.15, -0.28) -- (0.07, -0.03);
\node[orange!90!black, font=\scriptsize, anchor=west]
  at (0.95, -0.44) {laser};

\filldraw[red!80!black] (0, 0) circle (3pt);
\node[red!80!black, font=\scriptsize, align=center, anchor=east]
  at (-0.25, 0.45) {CBS IP\\($f_{\rm CBS}\le 10^{-6}$)};
\draw[gray!60, line width=0.3pt] (-0.22, 0.32) -- (-0.05, 0.08);

\draw[->, purple!60!black, line width=0.6pt]
  (0,0) arc[start angle=90, end angle=65, radius=8.5];
\draw[component_fill, fill=purple!25]
  (3.55, -1.02) rectangle (4.05, -0.62);
\node[purple!60!black, font=\tiny, align=center, anchor=north]
  at (3.8, -1.06) {$e^-$ absorber};
\node[purple!60!black, font=\tiny, anchor=west]
  at (1.60, -0.98) {scattered $e^-$};
\draw[gray!60, line width=0.3pt] (2.28, -0.86) -- (2.50, -0.35);

\fill[red!20, opacity=0.5] (0, 0) -- (5.5, 0.22) -- (5.5, -0.22) -- cycle;
\draw[arrow_gamma]
  (0, 0) -- (5.5, 0);
\node[red!70!black, font=\scriptsize, align=center] at (3.0, 0.78)
  {Backscattered $\gamma$ cone (100 m vacuum drift)};

\draw[component_fill, fill=gray!50]
  (5.5, -0.7) rectangle (6.1, 0.7);
\fill[white] (5.5, -0.1) rectangle (6.1, 0.1);
\draw[gray!70, line width=0.3pt]
  (5.65, -0.4) -- (5.95, -0.55);
\draw[gray!70, line width=0.3pt]
  (5.65, 0.4) -- (5.95, 0.55);

\node[font=\scriptsize, align=center] (lab_w) at (4.2, 2.4)
  {Primary W\\collimator};
\draw[gray!60, line width=0.3pt] (lab_w.south) -- (5.62, 0.72);

\draw[component_fill, fill=cyan!10]
  (6.4, -0.5) rectangle (7, 0.5);

\node[font=\scriptsize, align=center] (lab_dip) at (6.8, 2.4)
  {Sweeping\\dipole};
\draw[gray!60, line width=0.3pt] (lab_dip.south) -- (6.8, 0.5);

\draw[arrow_gamma, dashed, line width=0.5pt]
  (0, 0) -- (5.55, 0.38);

\filldraw[orange!80!black, fill=yellow!80!orange] (5.78, 0.38) circle (1.6pt);
\foreach \a in {20, 65, 110, 155, 200, 245, 290, 335}
  \draw[orange!80!black, line width=0.4pt]
    (5.78, 0.38) -- ++(\a:0.13);
\node[font=\tiny, anchor=south, align=center] at (5.92, 1.35)
  {$e^+e^-$ shower\\(in jaw)};
\draw[gray!60, line width=0.3pt] (5.90, 1.32) -- (5.80, 0.52);
\draw[->, teal!50!black, line width=0.4pt]
  (5.9, 0.34) .. controls (6.5, 0.28) and (6.95, 0.36) .. (7.5, 0.5);
\node[teal!50!black, font=\scriptsize, anchor=west]
  at (7.5, 0.55) {$e^-$};
\draw[->, olive!50!black, line width=0.4pt]
  (5.9, 0.30) .. controls (6.5, 0.12) and (6.95, -0.12) .. (7.5, -0.35);
\node[olive!50!black, font=\scriptsize, anchor=west]
  at (7.5, -0.35) {$e^+$};

\draw[component_fill, fill=gray!70]
  (7.8, -0.5) rectangle (8.2, 0.5);
\fill[white] (7.8, -0.08) rectangle (8.2, 0.08);

\node[font=\scriptsize, align=center] (lab_pb) at (8.7, 2.3)
  {Pb collimator};
\draw[gray!60, line width=0.3pt] (lab_pb.south) -- (8.0, 0.5);

\draw[gray, line width=0.4pt, dashed]
  (8.2, -0.3) -- (9.6, -0.3);
\draw[gray, line width=0.4pt, dashed]
  (8.2, 0.3) -- (9.6, 0.3);
\node[font=\scriptsize, align=center] at (8.9, -0.7) {Vacuum};

\fill[red!20, opacity=0.5] (0, 0) -- (5.5, 0.22) -- (5.5, -0.22) -- cycle;
\draw[arrow_gamma]
  (5.5, 0) -- (9.8, 0);

\draw[component_fill, fill=green!12]
  (9.8, -2.2) rectangle (12.4, 2.2);
  
\node[font=\scriptsize, align=center, anchor=south]
  at (11.11, 2.6) {Cylindrical detector};
\node[font=\scriptsize, align=center, anchor=south]
  at (11.11, 2.35) {+};
\node[font=\scriptsize, align=center, anchor=south]
  at (11.11, 2.1) {Pair Spectrometer};
  
\node[font=\scriptsize, align=center, anchor=north]
  at (11.11, -2.3) {event-level $E_\gamma$ reconstruction};

\draw[component_fill, fill=blue!10, dashed]
  (9.8, -0.7) rectangle (11.0, 0.7);
\node[font=\tiny, anchor=north]
  at (10.4, -0.22) {(cryostat)};
  
\draw[component_fill, fill=blue!30]
  (9.9, -0.2) rectangle (10.6, 0.2);
\node[font=\tiny, anchor=south]
  at (10.25, -0.22) {NH$_3$};
  
\fill[white]
  (10.6, -0.15) -- (12.4, -0.30) -- (12.4, 0.30) -- (10.6, 0.15) -- cycle;
\draw[gray!60, line width=0.3pt]
  (10.6, -0.15) -- (12.4, -0.30);
\draw[gray!60, line width=0.3pt]
  (10.6, 0.15) -- (12.4, 0.30);

\draw[arrow_gamma]
  (10.6, 0) -- (14.7, 0);

\node[font=\scriptsize, anchor=west, align=center]
  at (12.3, 0.14) {forward};
\node[font=\scriptsize, anchor=west, align=center]
  at (12.3, -0.12) {beam hole};

\draw[component_fill, fill=gray!80]
  (14.7, -1.1) rectangle (15.6, 1.1);
\draw[component_fill, fill=gray!50]
  (15.6, -1.2) rectangle (16.2, 1.2);
\draw[component_fill, fill=brown!40]
  (16.2, -1.3) rectangle (16.9, 1.3);
  
\node[font=\scriptsize, align=center, anchor=south]
  at (15.8, 1.4) {Beam dump};
\node[font=\scriptsize, align=center, anchor=north]
  at (15.8, -1.4) {W / Pb / concrete};
\node[font=\scriptsize, align=center, anchor=north]
  at (15.8, -1.8) {(water-cooled)};

\draw[line width=0.4pt] (-6, -3.2) -- (17, -3.2);
\foreach \x/\d in {0/{0}, 5.8/{100}, 6.8/{$\sim 101$}, 8.0/{$\sim 102$},
                   11.11/{$\sim 104$}, 15.8/{$\sim 110$}}
  {\draw[line width=0.2pt] (\x, -3.2) -- (\x, -3.3);
   \node[font=\tiny, anchor=north] at (\x, -3.3) {\d\,m};}

\node[font=\tiny, anchor=west, gray]
  at (-6.14, -3.) {Schematic layout (not to scale)};

\end{tikzpicture}

}
\caption{Schematic layout of the proposed CBS-based polarized gamma-ray facility at FCC-ee. The CBS interaction point is followed by a $100$~m vacuum drift, a primary tungsten collimator, a sweeping dipole magnet for $e^{+}e^{-}$ shower removal, a secondary lead collimator, the NH$_{3}$ DNP target surrounded by a cylindrical detector with an integrated pair spectrometer, and a downstream beam dump. The Compton-scattered electrons ($0.19$--$1.0\,E_{e}$) are separated in the first booster dipole and stopped in a local absorber; the sweeping dipole removes $e^{+}e^{-}$ pairs produced by the off-axis photon halo in the primary-collimator jaws. Not to scale.}
\label{fig:setup}
\end{figure*}

The proposed facility converts circularly polarized laser photons into high-energy, highly polarized gamma-ray beams via Compton backscattering off the FCC-ee electron beam and directs the resulting photon beam onto a polarized nuclear target for spin asymmetry measurements. A schematic view of the system is shown in Fig.~\ref{fig:setup}.

 The electron beam of the FCC-ee full-energy booster intersects a focused, counter-propagating laser pulse at a dedicated conversion point in a straight section. By inverse Compton scattering, laser photons are boosted to gamma-ray energies and emitted in the forward direction, while the scattered electrons, having lost up to 81\% of their energy, are separated from the nominal orbit in the first downstream arc dipole and intercepted by a local absorber; the unscattered electrons continue undisturbed around the ring.

The booster cycle structure~\cite{FCCFSRvol2,Agapov2026} defines two operating scenarios. In the \emph{parasitic} scenario, the laser fires only during the 0.1~s extraction flat-top of each top-up ramp, with a Compton fraction $f_{\rm CBS}=10^{-8}$ per bunch crossing; since an electron traverses the conversion point only for the duration of a single cycle, its cumulative scattering probability is $P_{\rm loss}=f_{\rm rev}\,t_{\rm ft}\,f_{\rm CBS}\simeq 3\times10^{-6}$, negligible for the top-up beam quality. In the \emph{dedicated} baseline scenario, extended-flat-top fills are inserted into the idle windows between top-up cycles (1.5--3~s of flat-top per top-up interval, depending on the operating mode); since this beam is not delivered to the collider, the Compton fraction can be raised to $f_{\rm CBS}=10^{-6}$, for a cumulative loss below 1\% per cycle. In both scenarios the collider luminosity is entirely unaffected, which is the key design choice that allows the photon-source program to coexist with the standard FCC-ee physics program.

The backscattered photon beam propagates through a $100$~m vacuum drift to a primary tungsten collimator, which serves as a radiation-containment and geometric-purification element rather than a polarization filter (see Sec.~\ref{sec:cbs_simulation}).  Off-axis photons of the beam halo impinging on the tungsten jaws produce electromagnetic showers inside the collimator body via $\gamma\to e^{+}e^{-}$ pair production and subsequent bremsstrahlung (the Compton-scattered electrons never reach the photon beamline, being removed magnetically inside the booster ring); the resulting $e^{+}e^{-}$ pairs are deflected out of the photon beam axis by a downstream sweeping dipole magnet (1--2~T over $\sim 1$~m), and the residual soft-photon shower component is absorbed by a secondary lead collimator, following standard photoproduction beamline practice at facilities such as GlueX at Jefferson Lab~\cite{GlueX2021}. The drift and collimator regions are operated under vacuum to avoid atmospheric pair-production losses and to maintain optical quality through the sweeping-magnet section.

Event-level photon-energy reconstruction and polarized event selection are performed in the detector region by a pair spectrometer (PS), again following the GlueX approach~\cite{GlueX2021}. The PS converts a small fraction of beam photons in a thin converter foil into $e^{+}e^{-}$ pairs whose energies are measured in a magnetic spectrometer, providing event-by-event reconstruction of $E_{\gamma}$ with a sub-percent relative resolution adequate for the band-edge selection (see Sec.~\ref{sec:cbs_simulation}).  In the parasitic scenario, $f_{\rm CBS}N_{e}\simeq 100$--$270$ photons enter the line per bunch crossing; with a 75~$\mu$m Be converter ($2\times10^{-4}\,X_{0}$) the PS records at most $\sim0.04$ $e^{+}e^{-}$ pairs per crossing, so individual conversions are unambiguous. In the dedicated scenario ($\sim10^{4}$ photons per crossing) the PS operates as a statistical monitor of the delivered spectrum and polarization; the event-level energy of an interacting photon is then reconstructed from the hadronic final state, with the PS providing the absolute energy-scale and flux calibration. Pairs produced in the NH$_{3}$ target itself ($\sim0.1\,X_{0}$) are emitted within $\sim1/\gamma_{e}$ of the beam axis and exit through the forward beam hole into the dump. Events with $E_{\gamma}>E_{\min}$, where $E_{\min}$ is chosen on the high-energy Compton edge such that the mean Stokes parameter satisfies $|\langle S_{2}\rangle|=0.90$ (the analytic limit for an unpolarized beam; Sec.~\ref{sec:cbs_simulation}), are kept for the spin-asymmetry analysis. This event-level selection plays the role that a polarization-filtering collimator could not (see Sec.~\ref{sec:cbs_simulation}). The polarized target is a solid NH$_{3}$ sample prepared by dynamic nuclear polarization and operated in a longitudinal field~\cite{CrabbMeyer1997}.

The non-interacting fraction of the photon beam carries a time-averaged power that ranges from $\sim2.4$~W in the $t\bar{t}$ mode to $\sim46$~W in the $WW$ mode in the dedicated baseline scenario (and below 0.1~W in the fully parasitic scenario) and must be safely absorbed downstream of the target without depositing energy in the detector. A forward beam hole through the detector, with angular acceptance of order $1$--$3^{\circ}$, allows the unscattered photons to pass into a tungsten/lead/concrete beam dump with active water cooling, following the design used for high-rate fixed-target photoproduction experiments~\cite{GlueX2021}. The corresponding geometric acceptance loss for forward open-charm events is quantified in Sec.~\ref{sec:asymmetry}.

The four FCC-ee operating modes provide electron beam energies of $45.6$, $80$, $120$, and $182.5$~GeV, yielding maximum backscattered photon energies of $37$, $65$, $98$, and $148$~GeV for the $Z$, $WW$, $ZH$, and $t\bar{t}$ modes respectively. The laser parameters and collimator design are determined in Secs.~\ref{sec:kinematics} and~\ref{sec:cbs_simulation}.

\section{CBS Kinematics and Laser Selection}
\label{sec:kinematics}

In the Compton backscattering process, low-energy circularly polarized laser photons with energy $\omega_{0}\sim\mathcal{O}(1\text{--}10$~eV$)$ are brought into head-on collision with ultra-relativistic electrons ($E_{e}\gg m_{e}$). By inverse Compton kinematics, the laser photon acquires a large fraction of the electron's energy and is scattered in the forward direction. When a circularly polarized laser is used (helicity $\lambda_{0}=\pm 1$), the polarization state is largely transferred to the backscattered photon, making it possible to achieve near-complete circular polarization at the highest scattered energies. This property is essential for spin asymmetry measurements, which require a well-defined photon helicity.

The kinematics of the process are governed by four-momentum conservation. For a head-on collision between an electron with four-momentum $p$ and a laser photon with four-momentum $k$, the squared center-of-mass energy in the laboratory frame is
\begin{equation}
s = (p+k)^{2} \simeq m_{e}^{2} + 2 p\cdot k = m_{e}^{2} + 4 E_{e}\,\omega_{0}.
\label{eq:s}
\end{equation}
The dimensionless kinematic parameter $\kappa$, which characterizes the energy-transfer capacity of the interaction, is defined as~\cite{Ginzburg1983,Telnov1990}
\begin{equation}
\kappa = \frac{s-m_{e}^{2}}{m_{e}^{2}} = \frac{4 E_{e}\omega_{0}}{m_{e}^{2}}.
\label{eq:kappa}
\end{equation}
The maximum energy attainable by the backscattered photon follows directly from momentum conservation,
\begin{equation}
\omega_{\max} = \frac{\kappa}{\kappa+1}\,E_{e}.
\label{eq:omega_max}
\end{equation}

Although increasing $\kappa$ yields higher photon energies, a fundamental upper limit arises at large $\kappa$. High-energy backscattered photons can interact with incoming laser photons in the conversion region via $\gamma+\gamma_{\rm laser}\to e^{+}e^{-}$, depleting the gamma-ray beam. The threshold condition for this process is $s_{\gamma\gamma}\geq 4 m_{e}^{2}$, where
\begin{equation}
s_{\gamma\gamma} = 4\,\omega_{\max}\omega_{0} \geq 4 m_{e}^{2}.
\label{eq:sgg}
\end{equation}
Substituting the expressions for $\omega_{\max}$ and $\omega_{0}$ in terms of $\kappa$,
\begin{equation}
\frac{\kappa}{\kappa+1}\,E_{e} \cdot \frac{\kappa\,m_{e}^{2}}{4 E_{e}}
\geq m_{e}^{2}
\;\Longrightarrow\;
\frac{\kappa^{2}}{\kappa+1}\geq 4,
\label{eq:kappa_quad}
\end{equation}
which reduces to $\kappa^{2}-4\kappa-4\geq 0$. The physically meaningful root gives
\begin{equation}
\kappa_{\max} = 2+2\sqrt{2}\simeq 4.83.
\label{eq:kappa_max}
\end{equation}

The facility design must therefore satisfy $\kappa\leq\kappa_{\max}$ to avoid pair-production losses. In practice, operating precisely at $\kappa_{\max}$ is inadvisable, as small fluctuations in beam energy or laser frequency could push the system into the pair-production regime. We therefore impose a 10\% safety margin,
\begin{equation}
\kappa\leq\kappa_{\rm safe}=0.9\,\kappa_{\max}\simeq 4.35.
\label{eq:kappa_safe}
\end{equation}

To maximize $\omega_{\max}$ in each operating mode independently, the laser photon energy $\omega_{0}$ is chosen for each mode to saturate this constraint, $\kappa=\kappa_{\rm safe}$. From Eq.~\eqref{eq:kappa},
\begin{equation}
\omega_{0} = \frac{\kappa_{\rm safe}\,m_{e}^{2}}{4 E_{e}}.
\label{eq:omega_0}
\end{equation}
The resulting kinematic parameters for each FCC-ee operating mode are summarized in Table~\ref{tab:kinematics}.

\begin{table}[t]
\centering
\caption{Kinematic parameters and laser wavelengths for each FCC-ee operating mode at $\kappa=\kappa_{\rm safe}=4.35$.}
\label{tab:kinematics}
\begin{tabular}{lcccc}
\toprule
FCC-ee mode & $\omega_{0}$ [eV] & $\lambda$ [nm] & $\kappa$ & $\omega_{\max}$ [GeV] \\
\midrule
$Z$         & 6.22 & 199.09 & 4.35 & 37.07  \\
$WW$        & 3.55 & 349.29 & 4.35 & 65.04  \\
$ZH$        & 2.36 & 523.93 & 4.35 & 97.56  \\
$t\bar{t}$  & 1.56 & 796.82 & 4.35 & 148.37 \\
\bottomrule
\end{tabular}
\end{table}

Since $\omega_{0}\propto 1/E_{e}$, the four operating modes require four distinct laser wavelengths spanning the deep ultraviolet to the near-infrared. For each value listed in Table~\ref{tab:kinematics}, a well-established laser technology exists at or near the required wavelength: a frequency-quadrupled Ti:Sapphire laser (fundamental near $797$~nm, fourth harmonic at $199$~nm) for the $Z$ mode, the third harmonic of a Nd:YLF laser ($350$~nm) for the $WW$ mode, a frequency-doubled Nd:YAG laser tuned near $524$~nm for the $ZH$ mode, and a Ti:Sapphire laser ($797$~nm) for the $t\bar{t}$ mode.

\begin{table}[t]
\centering
\caption{FCC-ee full-energy booster parameters at the CBS conversion point for each operating mode, from the FCC Feasibility Study Report, Vol.~2, Table~4.1~\cite{FCCFSRvol2}. $N_e^{\rm fil}$ ($N_e^{\rm top}$) is the bunch population for initial-filling (top-up) cycles; $n_b$ is the number of stored booster bunches; within each collider top-up interval $T_{\rm topup}$ the booster executes $n_{\rm ramp}$ ramps of 0.1~s flat-top each, leaving an idle window used for the dedicated CBS fills (Sec.~\ref{sec:setup}).}
\label{tab:booster_params}
\begin{tabular}{lcccc}
\toprule
Parameter & $Z$ & $WW$ & $ZH$ & $t\bar{t}$ \\
\midrule
Beam energy $E_e$ [GeV]                          & 45.6  & 80.0  & 120.0 & 182.5 \\
$N_e^{\rm fil}$ ($\times 10^{10}$)               & 2.725 & 1.268 & 1.268 & 1.268 \\
$N_e^{\rm top}$ ($\times 10^{10}$)               & 2.725 & 1.035 & 1.268 & 1.125 \\
Stored bunches $n_b$                             & 1120  & 928   & 300   & 64    \\
Ramps per top-up cycle $n_{\rm ramp}$            & 10    & 2     & 1     & 1     \\
Top-up interval $T_{\rm topup}$ [s]              & 43.4  & 14.8  & 11.3  & 10.4  \\
Horiz.\ emittance $\varepsilon_x$ [nm\,rad]      & 0.087 & 0.27  & 0.61  & 1.40  \\
Vert.\ emittance $\varepsilon_y$ [pm\,rad]       & 1.75  & 5.37  & 12.1  & 28.0  \\
RMS bunch length $\sigma_z$ [mm]                 & 2.43  & 2.56  & 2.26  & 1.98  \\
RMS energy spread $\sigma_E$ ($\times 10^{-3}$)  & 0.38  & 0.67  & 1.01  & 1.53  \\
\bottomrule
\end{tabular}
\end{table}

 The CBS conversion point is located in a dedicated straight section of the FCC-ee full-energy booster~\cite{FCCFSRvol2,Agapov2026}. The interaction-point optics follow directly from two observations. First, the backscattered energy spectrum and the polarization transfer $S_{2}(E_\gamma)$ depend only on $(E_{e},\omega_{0})$: the electron angular divergence tilts the emission cone but modifies the effective crossing angle (and hence $\kappa$) only at $\mathcal{O}(\sigma^{\prime 2}/4)\lesssim 10^{-11}$, while the Compton-edge sharpness is smeared only by the relative energy spread, $\delta\omega_{\max}/\omega_{\max}=\sigma_{E}(2+\kappa)/(1+\kappa)\simeq 0.05$--$0.18\%$, negligible against the accepted band width. The divergence therefore only inflates the photon-beam \emph{envelope} (collimator aperture and target spot). We keep the envelope radiation-dominated by saturating
\begin{equation}
\sigma^{\prime}_{x}=\sqrt{\varepsilon_{x}/\beta^{*}}=1/\gamma
\;\Longrightarrow\;
\beta^{*}=\varepsilon_{x}\gamma^{2},\qquad
\sigma_{x}=\varepsilon_{x}\gamma,
\label{eq:beta_xy}
\end{equation}
with the vertical plane following from the FSR emittances, $\sigma_{y}=\sqrt{\varepsilon_{y}\beta^{*}}$ (flat beam, $\varepsilon_{y}/\varepsilon_{x}=2\times10^{-2}$~\cite{FCCFSRvol2}). Second, the laser waist is chosen to minimise the overlap area, the only hard constraint being the hourglass effect: the Rayleigh length $z_{R}=4\pi\sigma_{L}^{2}/\lambda$ must cover the bunch, $z_{R}\geq 3\sigma_{z}$, giving $\sigma_{L}=\sqrt{3\sigma_{z}\lambda/4\pi}=11$--$19~\mu$m.

The laser pulse energy at full conversion, $E_{\rm pulse}^{(0)}$, is determined following Ref.~\cite{Alekhin1999}. The condition that every electron in the bunch encounters at least one laser photon requires
\begin{equation}
n_{0}\,\sigma_{C} = S_{\rm eff},
\label{eq:nsigma}
\end{equation}
 where $n_{0}$ is the number of laser photons per pulse, $\sigma_{C}$ is the total inverse-Compton cross section evaluated from the Klein--Nishina formula at $\kappa=\kappa_{\rm safe}$, and $S_{\rm eff}$ is the effective geometric overlap area of the Gaussian electron and laser transverse profiles,
\begin{equation}
S_{\rm eff}=2\pi\sqrt{(\sigma_{L}^{2}+\sigma_{x}^{2})(\sigma_{L}^{2}+\sigma_{y}^{2})},
\label{eq:Seff}
\end{equation}
which reduces to the familiar $4\pi\sigma^{2}$ in the round, matched limit $\sigma_{L}=\sigma_{x}=\sigma_{y}=\sigma$~\cite{Alekhin1999}. The Klein--Nishina cross section is~\cite{Ginzburg1983,Telnov1990}
\begin{equation}
\sigma_{C}(x) = \frac{3}{4}\,\sigma_{T}\!\left[
\frac{1+x}{x^{3}}\!\left(\frac{2 x(1+x)}{1+2x} - \ln(1+2x)\right)
+ \frac{\ln(1+2x)}{2x} - \frac{1+3x}{(1+2x)^{2}}\right],
\label{eq:KN}
\end{equation}
where $\sigma_{T}=(8/3)\pi r_{e}^{2}=6.65\times 10^{-29}$~m$^{2}$ is the Thomson cross section and $x=\kappa/2$. At $\kappa_{\rm safe}=4.35$, Eq.~\eqref{eq:KN} gives $\sigma_{C}=2.00\times 10^{-29}$~m$^{2}$.

Combining Eq.~\eqref{eq:nsigma} with $E_{\rm pulse}^{(0)}=n_{0}\,\omega_{0}$, the full-conversion pulse energy is
\begin{equation}
E_{\rm pulse}^{(0)} = \frac{S_{\rm eff}\,\omega_{0}}{\sigma_{C}}.
\label{eq:Epulse0}
\end{equation}
For parasitic operation that preserves the FCC-ee collider luminosity, the laser pulse energy is reduced by the Compton fraction $f_{\rm CBS}$:
\begin{equation}
E_{\rm pulse}^{\rm op} = f_{\rm CBS}\,E_{\rm pulse}^{(0)},
\label{eq:Epulse_op}
\end{equation}
where $f_{\rm CBS}=10^{-8}$ ($10^{-6}$) in the parasitic (dedicated) scenario of Sec.~\ref{sec:setup}. The resulting CBS conversion point parameters are summarized in Table~\ref{tab:cbs_point}.

\begin{table}[t]
\centering
\caption{CBS conversion-point design for each operating mode: flat-beam optics saturating $\sigma^{\prime}_{x}=1/\gamma$ [Eq.~\eqref{eq:beta_xy}] with the FSR emittances of Table~\ref{tab:booster_params}, laser waist from the hourglass condition $z_{R}=3\sigma_{z}$, and full-conversion pulse energy from Eq.~\eqref{eq:Seff}. Operational pulse energies $E_{\rm pulse}^{\rm op}=f_{\rm CBS}\,E_{\rm pulse}^{(0)}$ are given in Table~\ref{tab:collimator}.}
\label{tab:cbs_point}
\begin{tabular}{lccccc}
\toprule
Mode & $\beta^{*}$ [m] & $\sigma_{x}$ [$\mu$m] & $\sigma_{y}$ [$\mu$m] & $\sigma_{L}$ [$\mu$m] & $E_{\rm pulse}^{(0)}$ [J] \\
\midrule
$Z$         & 0.69 & 7.8  & 1.1  & 10.8 & 44.8 \\
$WW$        & 6.6  & 42.3 & 6.0  & 14.6 & 125.8 \\
$ZH$        & 33.6 & 143  & 20.2 & 16.8 & 450 \\
$t\bar{t}$  & 179  & 500  & 70.7 & 19.4 & 2867 \\
\bottomrule
\end{tabular}
\end{table}

The full-conversion pulse energies $E_{\rm pulse}^{(0)}$ lie in the 45~J--2.9~kJ range; at the operational Compton fractions these reduce to 0.4--29~$\mu$J (parasitic) and 45~$\mu$J--2.9~mJ (dedicated), with time-averaged laser powers during firing of 90--610~W in the dedicated scenario, within reach of cavity-enhanced or high-repetition amplifier systems at the corresponding wavelengths.

\section{CBS Photon Beam and Collimator Design}
\label{sec:cbs_simulation}

\begin{figure}[!htbp]
\centering
\includegraphics[width=\linewidth]{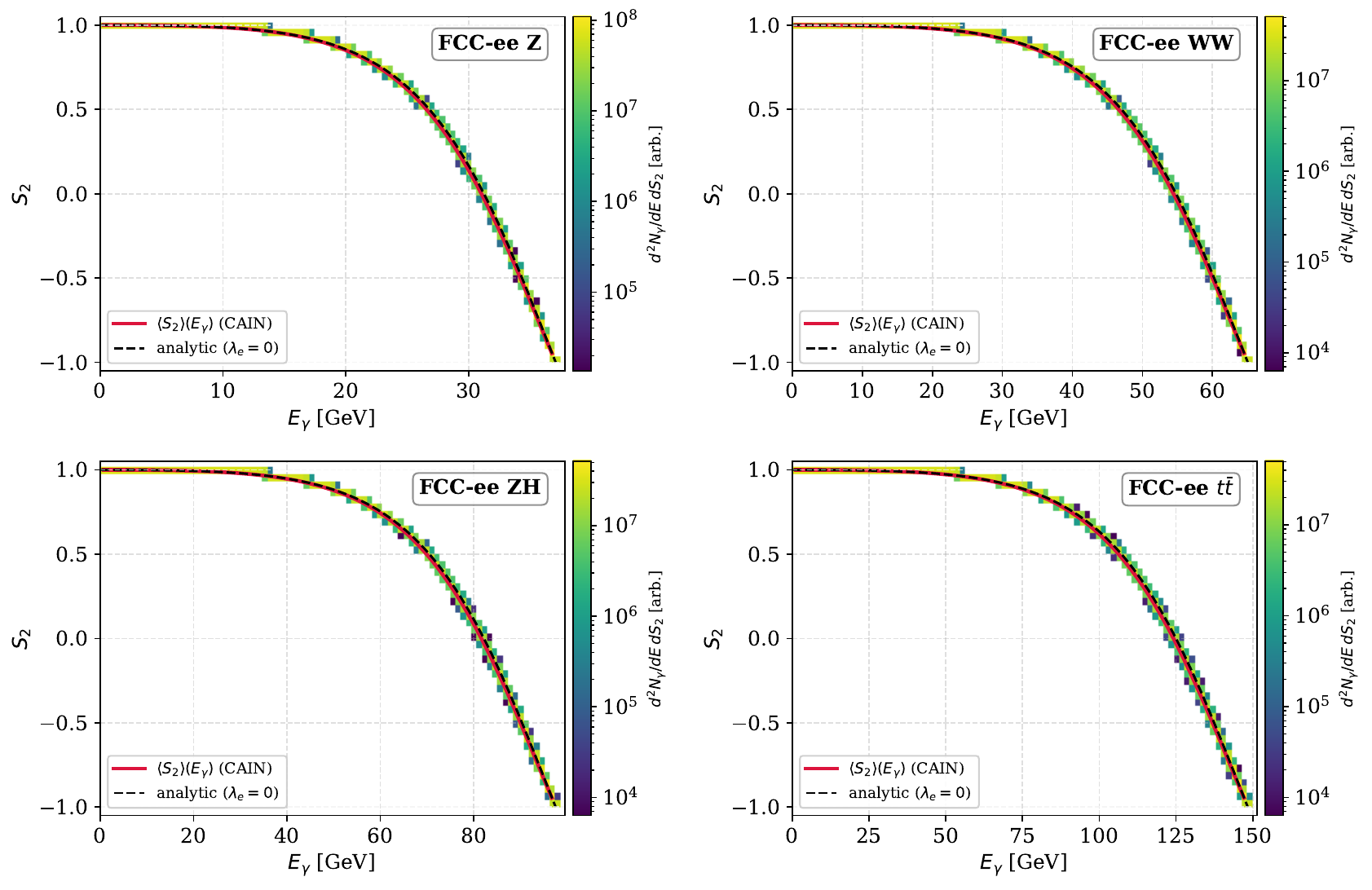}
\caption{Two-dimensional density of the Stokes circular-polarization parameter $S_{2}$ versus $E_{\gamma}$ (colour scale), with the CAIN weighted-mean profile $\langle S_{2}\rangle(E_{\gamma})$ (solid red) and the analytic transfer function of Eq.~\eqref{eq:S2transfer} for an unpolarized beam (dashed black). Low-energy photons retain the laser helicity ($\langle S_{2}\rangle\to+1$); Compton-edge photons acquire the opposite helicity ($\langle S_{2}\rangle\to-1$). The high-energy band $E_{\gamma}>E_{\min}$ is selected event-by-event by the pair spectrometer.}
\label{fig:polarization}
\end{figure}

 To characterise the backscattered photon beam in each FCC-ee operating mode, we performed Monte Carlo simulations with the CAIN code~\cite{CAIN1995}. The electron beam parameters are those of Table~\ref{tab:booster_params} with the interaction-point optics of Table~\ref{tab:cbs_point}; the beam is unpolarized, as appropriate for the booster (linac injection, no Sokolov--Ternov build-up within a few-second cycle). The laser is circularly polarized ($\lambda_{0}=+1$) with pulse length matched to the bunch, $\sigma_{z}^{\rm laser}=\sigma_{z}$. Since the spectral shape and $S_{2}(E_\gamma)$ are independent of the transverse overlap profile, the simulation employs a wide, dilute laser (waist $3\sigma_{x}$, on-axis scattering probability $P=0.3$) rather than the operational full-conversion pulse: this reproduces the strictly single-scattering operational regime ($P\sim f_{\rm CBS}\leq10^{-6}$) while retaining high statistics ($2\times10^{6}$ macro-electrons per mode). First-generation Compton photons are selected; the removed second-scatter component (a $\sim18\%$ sampling artefact of the enhanced $P$, kinematically confined below $0.103\,\omega_{\max}$) cannot populate the analysis band. The filtered spectra and polarization profiles agree with the analytic single-Compton expectation (Fig.~\ref{fig:polarization}); the operational photon flux is applied analytically below.

\begin{figure}[!htbp]
\centering
\includegraphics[width=\linewidth]{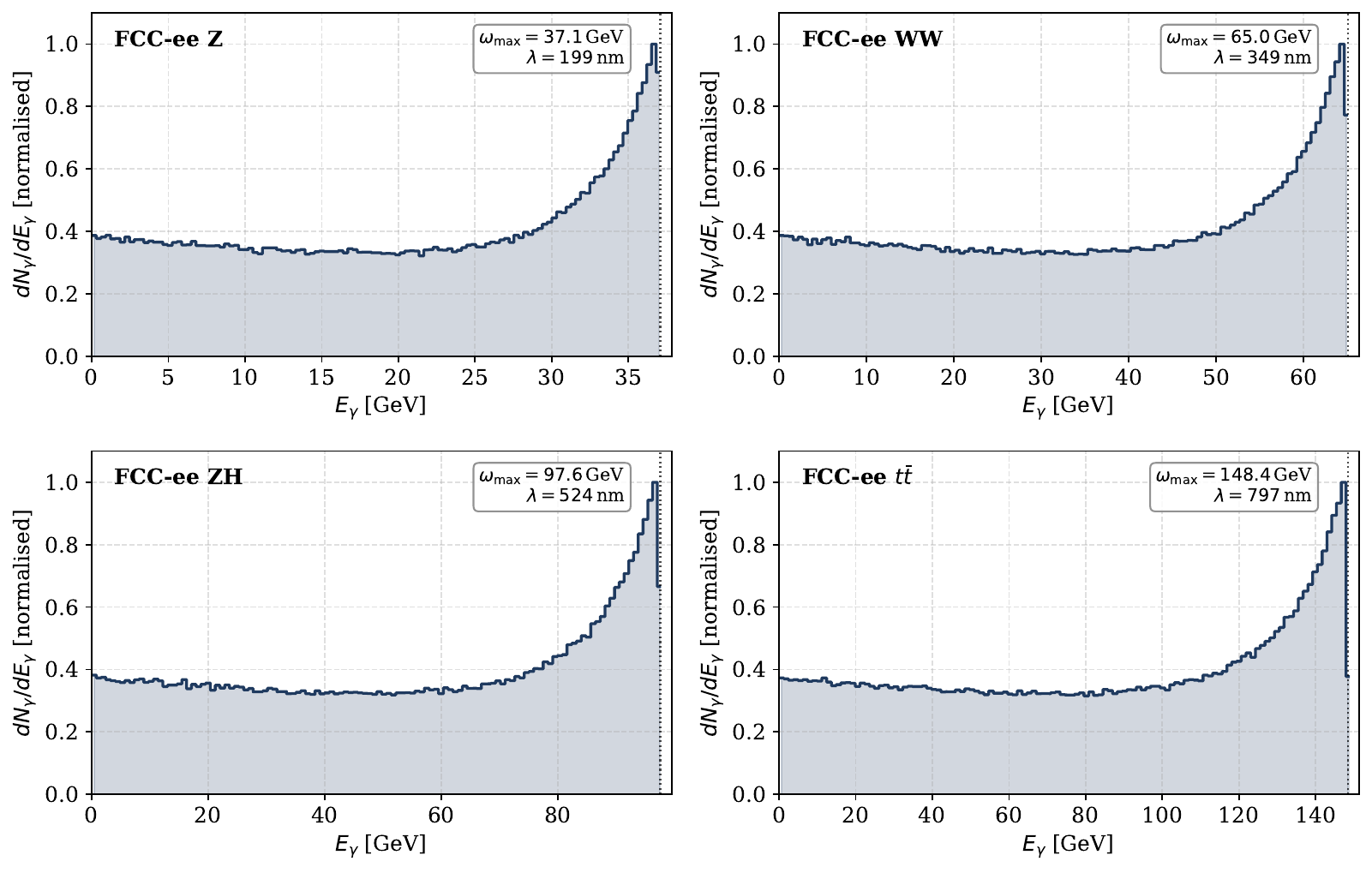}
\caption{Backscattered photon energy spectrum $dN_{\gamma}/dE_{\gamma}$ for the four FCC-ee operating modes, normalised to unity at the maximum. The vertical dashed line indicates the kinematic edge $\omega_{\max}$. The spectra follow the ideal Klein--Nishina shape for $\kappa=4.35$ for the unpolarized booster beam.}
\label{fig:spectra}
\end{figure}

\begin{figure}[!htbp]
\centering
\includegraphics[width=\linewidth]{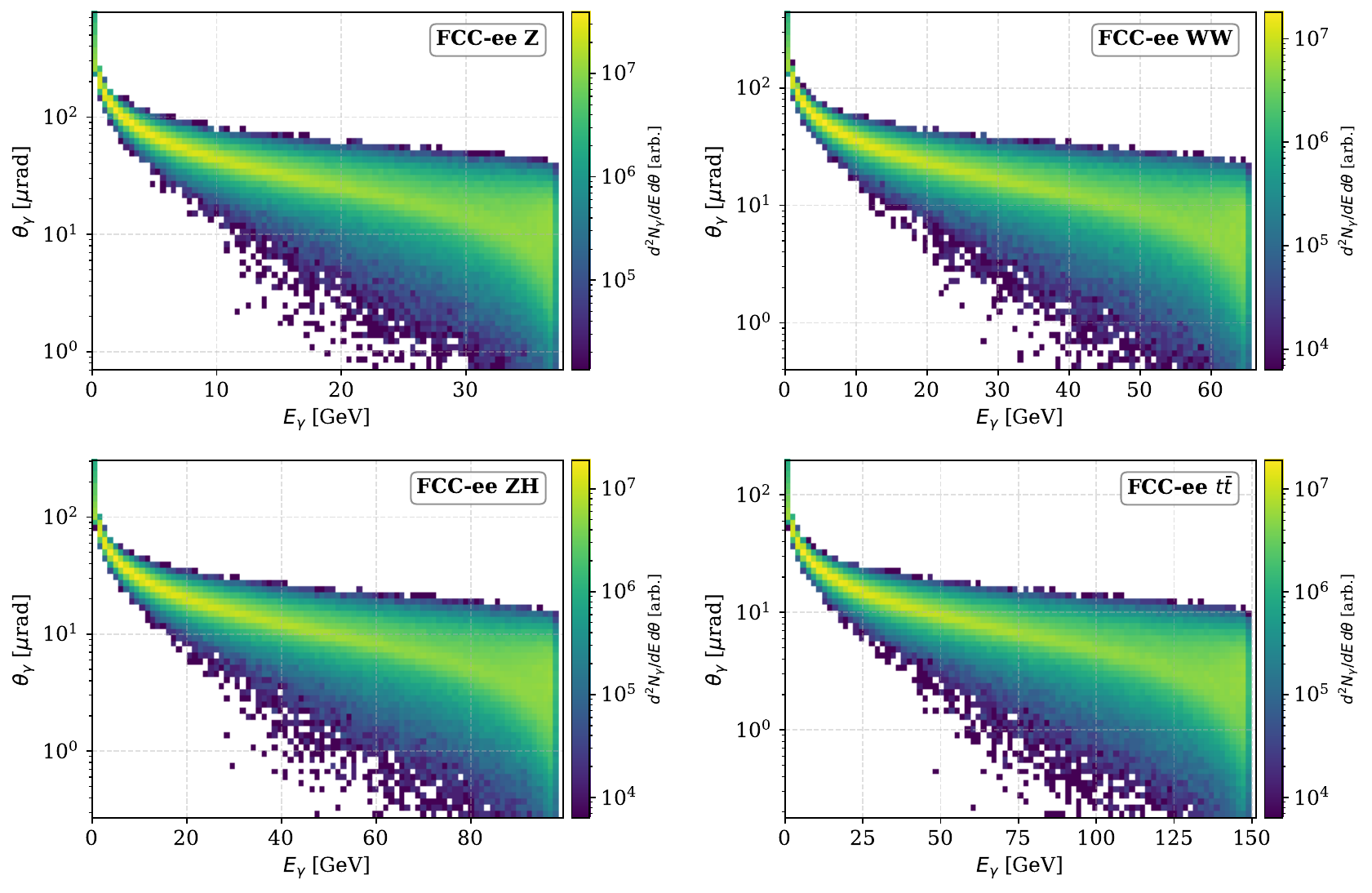}
\caption{Photon scattering angle $\theta_{\gamma}$ versus $E_{\gamma}$ (logarithmic scale). The 99\% containment angle $\theta_{99}$ sets the geometric requirement on the vacuum aperture and the collimator radius. The large angles at low $E_{\gamma}$ are of kinematic origin ($\theta_\gamma \propto 1/E_\gamma^{1/2}$) and not due to beam emittance, which is held at $\sigma^{\prime}_{x}=1/\gamma=2.8$--$11.2~\mu$rad by the interaction-point design.}
\label{fig:theta_E}
\end{figure}

Figure~\ref{fig:spectra} shows the energy spectrum $dN_{\gamma}/dE_{\gamma}$ of the backscattered photons in each operating mode. The electron angular divergence is set to $\sigma^{\prime}_{x}=1/\gamma$ by design [Eq.~\eqref{eq:beta_xy}], so the spectrum retains the ideal Klein--Nishina shape while the photon-beam envelope stays radiation-dominated. The resulting spectra follow the ideal Klein--Nishina shape for $\kappa=4.35$: they peak at low photon energy and exhibit a characteristic upturn toward the Compton edge at $E_{\gamma}=\omega_{\max}$, where the differential cross section has a secondary maximum~\cite{Ginzburg1983,Telnov1990}. Figure~\ref{fig:polarization} shows the circular polarization Stokes parameter $S_{2}$ as a function of $E_{\gamma}$.  For an unpolarized electron beam and a fully circular laser, the mean Stokes parameter of the scattered photon is given by the Compton polarization-transfer function~\cite{Ginzburg1983}
\begin{equation}
\langle S_{2}\rangle(y)=
\lambda_{0}\,\frac{(1-2r)\left[(1-y)^{-1}+1-y\right]}
{(1-y)^{-1}+1-y-4r(1-r)},\qquad
r=\frac{y}{\kappa(1-y)},
\label{eq:S2transfer}
\end{equation}
with $y=E_{\gamma}/E_{e}$: low-energy photons retain the laser helicity ($\langle S_{2}\rangle\to+1$; the simulated band average below $0.05\,\omega_{\max}$ is $+0.9999$), the transfer reverses sign at intermediate energies, and $\langle S_{2}\rangle\to-1$ exactly at the Compton edge. At fixed $E_{\gamma}$ the circular component is single-valued; the CAIN profile reproduces Eq.~\eqref{eq:S2transfer} across the full spectrum (Fig.~\ref{fig:polarization}). The angle--energy correlation is shown in Fig.~\ref{fig:theta_E}, where $\theta_{\gamma}$ is the polar angle of the backscattered photon with respect to the electron beam axis: high-energy photons populate a narrow forward cone, while lower-energy photons fill a wider angular distribution, as expected from CBS kinematics ($\theta_{\gamma}\propto 1/E_{\gamma}^{1/2}$ in the limit $\theta\gg 1/\gamma$). The weighted RMS spot sizes $\sigma_{x},\sigma_{y}$ at the conversion point, the angular RMS divergence $\sigma_{\theta_{x}},\sigma_{\theta_{y}}$, the 99\% containment angle $\theta_{99}$ extracted from the simulated photon distribution, and the Compton-edge polarization within $E_{\gamma}>0.95\,\omega_{\max}$ are summarized in Table~\ref{tab:precoll} (the operational PS cut at $E_{\gamma}>E_{\min}$ yields $f_{\rm edge}\simeq 7\%$, as detailed in Table~\ref{tab:collimator}).

\begin{table*}[t]
\centering
\caption{Pre-collimation photon-beam characteristics from the CAIN simulations (first-scatter photons). The source spot sizes are the design values $\sigma_{x},\sigma_{y}$ of Table~\ref{tab:cbs_point}; the scattering-vertex RMS observed in the simulation equals the beam--laser product-Gaussian, $0.95\,\sigma_{x}$, as expected. The Compton-edge band here is the diagnostic $E_{\gamma}>0.95\,\omega_{\max}$; the analysis band is defined in Table~\ref{tab:collimator}.}
\label{tab:precoll}
\begin{tabular}{lccccccc}
\toprule
Mode & $\sigma_{x}$ [$\mu$m] & $\sigma_{y}$ [$\mu$m] &
$\sigma_{\theta_{x}}$ [$\mu$rad] & $\sigma_{\theta_{y}}$ [$\mu$rad] &
$\theta_{99}$ [$\mu$rad] & $\langle |S_{2}|\rangle_{\rm edge}$ &
Edge fraction [\%] \\
\midrule
$Z$         & 7.8 & 1.1  & 61.4 & 57.5 & 249.4 & 0.848 & 11.1 \\
$WW$        & 42.3 & 6.0 & 37.3 & 32.3 & 143.6 & 0.848 & 11.1 \\
$ZH$        & 143 & 20.2 & 25.6 & 22.7 &  95.0 & 0.848 & 11.1 \\
$t\bar{t}$  & 500 & 70.7 & 15.1 & 15.1 &  62.3 & 0.848 & 11.1 \\
\bottomrule
\end{tabular}
\end{table*}

\subsection{Role of the collimator: radiation envelope, not polarization filter}

A diagnostic study of the CAIN photon sample shows that a purely geometric angular cut, even at the 99\% containment angle of the photon population, does \emph{not} produce a strongly polarized beam: the mean Stokes parameter of all photons passing such a cut remains far from the analysis requirement ($|\langle S_{2}\rangle|\ll0.9$) in every operating mode. The reason is that, in the booster configuration with large $\beta^{*}$, the CBS spectrum follows the ideal Klein--Nishina distribution which strongly favors low-energy photon production. Low- and intermediate-energy photons, which carry $S_{2}\simeq 0$, are emitted at large angles but still pass through any realistically-sized collimator aperture. A collimator selecting on $\theta$ alone therefore cannot deliver a polarized beam.

The collimator system is consequently retained for its remaining design functions: (i) radiation containment, terminating the photon halo at the design vacuum aperture; (ii) geometric definition of the beam envelope; and (iii) defining the beam dump entrance. The primary tungsten collimator is realised as a bulk W block of thickness $t_{W}\geq 6.2$~cm, sufficient to suppress off-axis photon transmission to the $10^{-6}$ level via pair production ($\lambda_{\rm pair}=(9/7) X_{0} =4.5$~mm in tungsten)~\cite{PDG2024,GlueX2021}. Electromagnetic shower products escaping the collimator aperture are removed by a downstream sweeping dipole magnet, which deflects the $e^{+}e^{-}$ pairs out of the beam axis, and a secondary lead collimator absorbs the residual soft-photon component. The geometric collimator aperture is set by the 99\% containment angle of the photon distribution,
\begin{equation}
d_{\rm col} = 2\, L\,\theta_{99},
\label{eq:dcol}
\end{equation}
with $L=100$~m. The resulting collimator diameters range from $d_{\rm col}=12.5$~mm in the $t\bar{t}$ mode to $49.9$~mm in the $Z$ mode (Table~\ref{tab:collimator}), accommodated by a correspondingly sized vacuum aperture in the drift section.

The drift section and collimator system are operated under vacuum: in atmospheric air, photons in the $30$--$150$~GeV range would suffer $\sim 23\%$ pair-production losses over a $100$~m drift, whereas in a residual vacuum of $\sim 5$~Torr the loss is below $0.2\%$. The vacuum extends through the sweeping dipole and the secondary collimator and is terminated at a thin beryllium window immediately upstream of the target region, with negligible loss ($\sim 0.1\%$ for a 500~$\mu$m Be foil).

\subsection{Event-level polarized photon selection by the pair spectrometer}
Because the collimator cannot itself deliver a polarized beam, the polarized photon selection is performed event-by-event in the detector by a pair spectrometer~\cite{GlueX2021}. The PS converts a small fraction of the incident photon beam in a thin converter foil and reconstructs the photon energy from the measured $e^{+}e^{-}$ pair with a resolution of $\sim 0.1\%$ at GlueX operating energies.  For the unpolarized booster beam, Eq.~\eqref{eq:S2transfer} fixes the achievable band-averaged polarization: $\langle S_{2}\rangle=-1$ is reached only exactly at the edge, and the average over any finite band is smaller in magnitude (e.g.\ $-0.85$ over the top 5\% of the spectrum). We therefore set the selection at $\langle S_{2}\rangle=-0.90$, which the trade-off analysis (Fig.~\ref{fig:tradeoff}) places at $E_{\min}/\omega_{\max}\simeq0.97$ with an accepted fraction $f_{\rm edge}\simeq7\%$ in all four modes (Table~\ref{tab:collimator}); the resulting effective beam polarization $P_{\gamma}=|\langle S_{2}\rangle_{\rm band}|=0.90$ enters the sensitivity projections of Sec.~\ref{sec:asymmetry}. Since $P_{\gamma}$ is computed from Eq.~\eqref{eq:S2transfer} and the measured spectrum, its uncertainty is dominated by the absolute energy-scale calibration of the band edge; a few-per-mil calibration, as achieved by the GlueX pair spectrometer~\cite{GlueX2021}, keeps it within the $\delta P_{\gamma}/P_{\gamma}=1\%$ polarimetry budget.

\subsection{Operational photon flux and parasitic-mode laser energy}
The annual integrated flux of polarized photons useful for the spin-asymmetry analysis is
\begin{equation}
\Phi_{\gamma}^{\rm eff} = f_{\rm CBS}\,N_{e}\,n_{b}\,f_{\rm rev}\,f_{\rm edge}\,f_{\rm duty}\,\Delta T,
\label{eq:flux}
\end{equation}
 where $N_{e}$ and $n_{b}$ are the booster bunch population and bunch count of Table~\ref{tab:booster_params} (top-up column in the parasitic scenario, filling column in the dedicated one), $f_{\rm rev}=3306.8$~Hz the revolution frequency of the 90.66~km ring~\cite{FCCFSRvol2}, $f_{\rm edge}$ the PS-selected band fraction, $f_{\rm duty}$ the fraction of the $\Delta T=10^{7}$~s/yr operating time spent at flat-top energy with the laser firing, and $f_{\rm CBS}$ as defined in Sec.~\ref{sec:setup}. In the parasitic scenario $f_{\rm duty}=n_{\rm ramp}t_{\rm ft}/T_{\rm topup}=0.9$--$2.3\%$. In the dedicated baseline, the hold time within each idle window is maximised by adjusting the number of filled bunches (the fill time scales with the bunch count), giving effective duty factors of $1.8\%$ ($Z$, 593 bunches, 1.5~s hold), $20.2\%$ ($WW$), $6.8\%$ ($ZH$) and $6.6\%$ ($t\bar{t}$).

\begin{figure}[!htbp]
\centering
\includegraphics[width=0.6\linewidth]{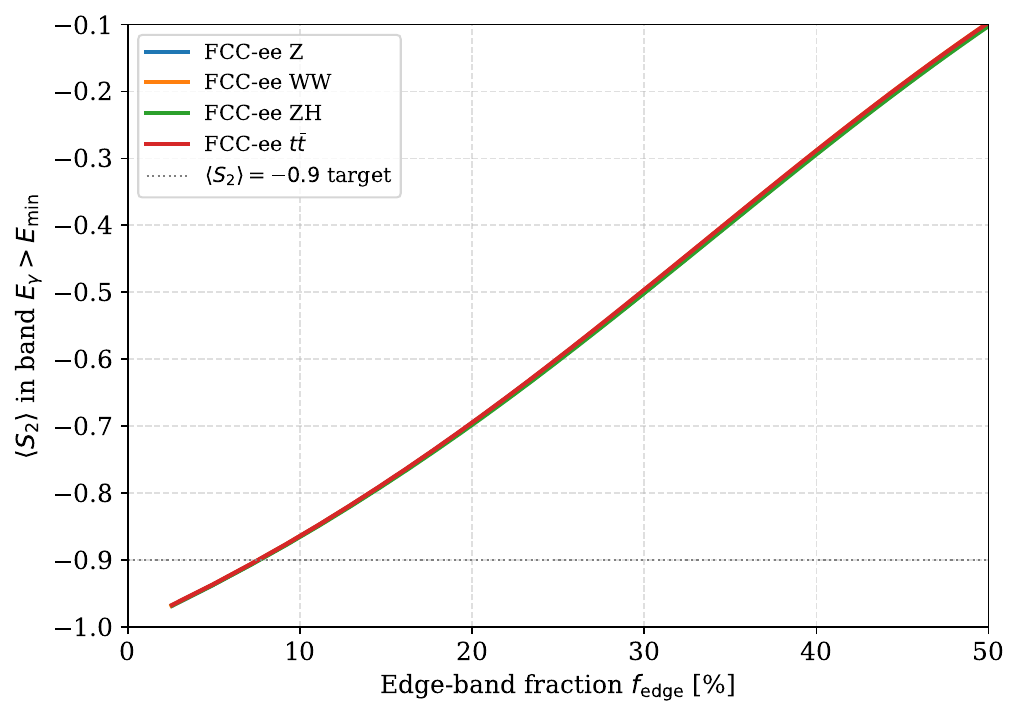}
\caption{Mean circular polarization $\langle S_{2}\rangle$ versus the fraction $f_{\rm edge}$ of photons in the high-energy Compton-edge band selected by the pair spectrometer cut $E_{\gamma}>E_{\min}$. The trade-off curve is universal in $\kappa$ and hence identical for all four modes; the working point $\langle S_{2}\rangle=-0.90$ accepts $f_{\rm edge}\simeq7\%$.}
\label{fig:tradeoff}
\end{figure}

The trade-off between $f_{\rm edge}$ and the achievable mean polarization $\langle S_{2}\rangle$, obtained by varying the lower edge $E_{\min}/\omega_{\max}$ of the accepted band, is shown in Fig.~\ref{fig:tradeoff}: the trade-off curve is universal in $\kappa$ and hence identical for all four modes; the working point $\langle S_{2}\rangle=-0.90$ accepts $f_{\rm edge}\simeq7\%$. The complete collimator design and operational parameters, including $E_{\rm pulse}^{\rm op}=f_{\rm CBS}\,E_{\rm pulse}^{(0)}$ in the millijoule range, are summarized in Table~\ref{tab:collimator}.

\begin{table*}[t]
\centering
\caption{Collimator design, edge-band selection ($\langle S_{2}\rangle=-0.90$) and annual polarized photon flux at $L=100$~m for the two operating scenarios of Sec.~\ref{sec:setup}.}
\label{tab:collimator}
\begin{tabular}{lcccccc|cc|cc}
\toprule
 & & & & & & &
\multicolumn{2}{c|}{Parasitic ($f_{\rm CBS}=10^{-8}$)} &
\multicolumn{2}{c}{Dedicated ($f_{\rm CBS}=10^{-6}$)} \\
Mode & $\theta_{99}$ [$\mu$rad] & $d_{\rm col}$ [mm] &
$E_{\min}$ [GeV] & $\tfrac{E_{\min}}{\omega_{\max}}$ & $f_{\rm edge}$ [\%] & $P_{\gamma}$ &
$\Phi_{\gamma}^{\rm eff}$ [yr$^{-1}$] & $E_{\rm pulse}^{\rm op}$ &
$\Phi_{\gamma}^{\rm eff}$ [yr$^{-1}$] & $E_{\rm pulse}^{\rm op}$ [mJ] \\
\midrule
$Z$        & 249 & 49.9 &  35.9 & 0.968 & 7.56 & 0.901 & $1.8\times10^{13}$ & 0.45~$\mu$J & $1.38\times10^{15}$ & 0.045 \\
$WW$       & 144 & 28.7 &  63.0 & 0.968 & 7.58 & 0.900 & $3.3\times10^{12}$ & 1.3~$\mu$J  & $5.97\times10^{15}$ & 0.126 \\
$ZH$       &  95 & 19.0 &  94.4 & 0.968 & 7.63 & 0.900 & $8.5\times10^{11}$ & 4.5~$\mu$J  & $6.50\times10^{14}$ & 0.450 \\
$t\bar{t}$ &  62 & 12.5 & 144.0 & 0.970 & 7.05 & 0.907 & $1.6\times10^{11}$ & 29~$\mu$J   & $1.25\times10^{14}$ & 2.87 \\
\bottomrule
\end{tabular}
\end{table*}

The yearly integrated polarized photon flux at the target in the dedicated baseline ranges from $1.25\times10^{14}$ in the $t\bar{t}$ mode to $5.97\times10^{15}$ in the $WW$ mode, one to two orders of magnitude above the fully parasitic option; as shown in Sec.~\ref{sec:asymmetry}, even the parasitic fluxes keep the measurement close to its systematic floor, so the choice of scenario affects operational flexibility more than the ultimate precision. The mean polarization of the delivered beam in the PS-selected band is uniformly $P_{\gamma}=0.90$, which enters the spin-asymmetry analysis of Sec.~\ref{sec:asymmetry}.

\section{Open Charm Photoproduction and Spin Asymmetry Projection}
\label{sec:asymmetry}

The pair-spectrometer-based edge-band selection of Sec.~\ref{sec:cbs_simulation} ensures that only photons in the high-energy Compton-edge band ($E_\gamma>E_{\min}$, $P_{\gamma}=0.90$) enter the spin-asymmetry analysis, with the low-$E_{\gamma}$ tail removed event-by-event by the PS reconstruction. The polarized gamma beam is directed onto a longitudinally polarized fixed target, where it interacts via open charm photoproduction $\gamma p\to c\bar{c}X$. At leading order in QCD this reaction proceeds exclusively through the photon--gluon fusion (PGF) subprocess $\gamma g\to c\bar{c}$, with no competing partonic mechanism, so that the spin-dependent observables provide direct sensitivity to the polarized gluon distribution $\Delta g(x)$ in the proton~\cite{BojakStratmann1999}. This same selection underlies the COMPASS open-charm program at CERN, which has used $\gamma^{*}p\to c\bar{c}X$ in muoproduction to extract $\Delta g/g$ in the kinematic range $x\sim 10^{-1}$~\cite{COMPASS2013}.

The unpolarized and polarized partonic cross sections at LO are~\cite{BojakStratmann1999}
\begin{widetext}
\begin{align}
\hat{\sigma}_{\gamma g\to c\bar{c}}(\hat{s}) &=
\frac{\pi\alpha\,\alpha_{s}(\mu^{2})\,e_{c}^{2}}{\hat{s}}
\left[(2+2\rho-\rho^{2})\ln\!\frac{1+\beta}{1-\beta}-2\beta(1+\rho)\right],
\label{eq:sigma_hat} \\[4pt]
\Delta\hat{\sigma}_{\gamma g\to c\bar{c}}(\hat{s}) &=
\frac{\pi\alpha\,\alpha_{s}(\mu^{2})\,e_{c}^{2}}{\hat{s}}
\left[(2-\rho)\ln\!\frac{1+\beta}{1-\beta}-2\beta\right],
\label{eq:dsigma_hat}
\end{align}
\end{widetext}
with $\rho=4 m_{c}^{2}/\hat{s}$ and $\beta=\sqrt{1-\rho}$. The charm-quark mass is fixed at the PDG value $m_{c}=1.27$~GeV~\cite{PDG2024} and the factorization and renormalization scales are set to $\mu^{2}=4 m_{c}^{2}$. The strong coupling $\alpha_{s}(\mu^{2})$ is evaluated at one loop with $\Lambda_{\rm QCD}=0.2$~GeV and $n_{f}=4$ active flavors above the charm threshold.

The hadronic cross sections are obtained by convoluting the partonic kernels with the gluon parton distribution function (PDF) of the proton,
\begin{align}
\sigma(\gamma p\to c\bar{c}X) &=
\int_{x_{\min}}^{1}\!dx\;g(x,\mu^{2})\,
\hat{\sigma}_{\gamma g\to c\bar{c}}(x s),
\label{eq:sigma_total}\\
\Delta\sigma(\gamma p\to c\bar{c}X) &=
\int_{x_{\min}}^{1}\!dx\;\Delta g(x,\mu^{2})\,
\Delta\hat{\sigma}_{\gamma g\to c\bar{c}}(x s),
\label{eq:dsigma_total}
\end{align}
where $s=2 m_{p} E_{\gamma}+m_{p}^{2}$ is the photon--proton center-of-mass energy squared in the laboratory frame, and $x_{\min}=4 m_{c}^{2}/s$. Both gluon distributions are accessed through the LHAPDF interface~\cite{LHAPDF2015}. The unpolarized $g(x,\mu^{2})$ is taken from the CT18 NLO global analysis~\cite{CT18}, and the polarized $\Delta g(x,\mu^{2})$ from the NNPDFpol2.0 NLO fit with missing-higher-order uncertainties~\cite{NNPDFpol20}, the most recent polarized PDF set that incorporates post-2017 RHIC pp constraints (STAR jets/dijets, PHENIX)~\cite{STAR2015,PHENIX2014}.

The leading-order photon-gluon fusion approximation receives substantial NLO QCD corrections, computed in Ref.~\cite{BojakStratmann1999}. For the kinematic range relevant to FCC-ee, $30~\text{GeV} < E_\gamma < 150~\text{GeV}$ at $\mu=2 m_c$, the NLO/LO $K$-factors take typical values $K_\sigma=\sigma^{\rm NLO}/\sigma^{\rm LO}\simeq 1.4$ for the unpolarized cross section and $K_{\Delta\sigma}=\Delta\sigma^{\rm NLO}/\Delta\sigma^{\rm LO}\simeq 1.2$ for the helicity-difference cross section, with a scale variation of $\pm 15\%$ under $\mu/2 < \mu < 2\mu$ absorbed in the multiplicative systematic budget below. Throughout this section we apply these $K$-factors to the LO results of Eqs.~\eqref{eq:sigma_total}--\eqref{eq:dsigma_total}, giving an effective NLO asymmetry $A_{LL}^{\rm NLO}=(K_{\Delta\sigma}/K_\sigma)\, A_{LL}^{\rm LO}\simeq 0.86\,A_{LL}^{\rm LO}$.

Experimentally, the longitudinal double-spin asymmetry is defined as
\begin{equation}
A_{LL}^{\rm exp} =
\frac{N^{\rightrightarrows}-N^{\rightleftarrows}}
     {N^{\rightrightarrows}+N^{\rightleftarrows}},
\label{eq:ALL_exp}
\end{equation}
where $N^{\rightrightarrows}$ ($N^{\rightleftarrows}$) is the number of events recorded with the photon and target helicities aligned (anti-aligned). In practice, the laser circular polarization is reversed on a pulse-by-pulse basis using a Pockels cell, so that both helicity configurations are accumulated within the same operating period and on the same target. The corresponding theoretical prediction is
\begin{equation}
A_{LL} = \frac{\Delta\sigma}{\sigma},
\label{eq:ALL_theory}
\end{equation}
obtained from Eqs.~\eqref{eq:sigma_total} and~\eqref{eq:dsigma_total}.

The polarized target is taken to be solid ammonia (NH$_{3}$) prepared by dynamic nuclear polarization (DNP) at $\sim 100$~mK in a $\sim 5$~T longitudinal field with $\sim 140$~GHz microwave irradiation, the standard configuration of COMPASS-type frozen-spin targets and of the SpinQuest experiment at Fermilab~\cite{CrabbMeyer1997}. The relevant target parameters used in the projection are a density $\rho_{\rm NH_{3}}=0.85$~g/cm$^{3}$, a thickness $t_{\rm tgt}=5$~cm, three polarizable hydrogen atoms per NH$_{3}$ molecule, an achievable proton polarization $P_{\rm t}=0.85$, and a dilution factor $f_{d}=3/M_{\rm NH_{3}}=0.176$.

The annual yield of polarized open-charm events on this target follows directly from the photon flux of Sec.~\ref{sec:cbs_simulation},
\begin{equation}
N_{\rm event} = \Phi_{\gamma}^{\rm eff}\,\sigma\,
N_{N/{\rm cm}^{2}}\,\varepsilon_{\rm geom},
\label{eq:Nevent}
\end{equation}
with $N_{N/{\rm cm}^{2}}=(\rho_{\rm NH_{3}}\,t_{\rm tgt}/M_{\rm NH_{3}})\, N_{A}\times 17\simeq 2.6\times 10^{24}$ nucleons per cm$^{2}$ in the target, and $\varepsilon_{\rm geom}=0.85$ a conservative geometric acceptance factor that accounts for the loss of forward open-charm events through the detector beam hole (Sec.~\ref{sec:setup}). Since forward and central charm events share the same partonic origin ($\gamma g\to c\bar{c}$), this acceptance loss affects only the event statistics and not the value of $A_{LL}$ itself. The dilution factor $f_{d}$ accounts for the fraction of polarizable protons (the three hydrogen atoms in NH$_{3}$), entering the statistical-uncertainty formula
\begin{equation}
\delta A_{LL}^{\rm stat} =
\frac{1}{P_{\gamma}\,P_{\rm t}\,f_{d}\,\sqrt{N_{\rm event}}},
\label{eq:dA_stat}
\end{equation}  with $P_{\gamma}=0.90$ the band-averaged beam polarization of Table~\ref{tab:collimator},
while the systematic uncertainty contains a multiplicative component, scaling with the asymmetry through the polarimetry and dilution-factor uncertainties, and an absolute (additive) component arising from background subtraction in the charm-tagging procedure,
\begin{equation}
\delta A_{LL}^{\rm syst} =
\sqrt{(A_{LL}\,\sigma_{\rm mult})^{2}+\sigma_{\rm add}^{2}},
\label{eq:dA_syst}
\end{equation}
with $\sigma_{\rm mult}$ obtained from $\delta P_{\gamma}/P_{\gamma}=1\%$ (CBS luminosity polarimetry), $\delta P_{\rm t}/P_{\rm t}=3\%$ (NMR readout) and $\delta f_{d}/f_{d}=2\%$ (target stoichiometry), giving $\sigma_{\rm mult}\simeq 3.7\%$, and $\sigma_{\rm add}=3\times 10^{-3}$ for the absolute background-subtraction uncertainty. The combined experimental uncertainty is therefore $\delta A_{LL}^{\rm exp}=\sqrt{(\delta A_{LL}^{\rm stat})^{2}+(\delta A_{LL}^{\rm syst})^{2}}$. The additive background-subtraction uncertainty $\sigma_{\rm add}=3\times 10^{-3}$ corresponds to a factor of $\sim 30$ reduction relative to the COMPASS open-charm analysis~\cite{COMPASS2013}, which obtained a total systematic uncertainty $\delta A_{LL}\sim 0.09$. The improvement reflects two distinct effects: (i) at leading order the photoproduction channel $\gamma g\to c\bar{c}$ proceeds with no inclusive deep-inelastic background, in contrast to the muon-DIS measurement of COMPASS, reducing the combinatorial background uncertainty by roughly an order of magnitude; and (ii) the CBS-based photon polarimetry with Pockels-cell helicity switching provides a factor of $\sim 3$ tighter beam-polarization control than the COMPASS muon-beam polarimetry. We further note that the unpolarized $g(x,\mu^2)$ from CT18 NLO is constrained by global DIS and LHC data to the percent level~\cite{CT18} and contributes negligibly compared to the $\sim 10$--$20\%$ uncertainty on $\Delta g(x,\mu^2)$ from NNPDFpol2.0; we therefore propagate only the polarized PDF uncertainty through the NNPDFpol2.0 replicas.

In addition to the experimental uncertainty, the theoretical prediction of $A_{LL}$ inherits an uncertainty from the polarized parton distributions. We compute it directly by evaluating $A_{LL}$ over the $100$ Monte Carlo replicas of NNPDFpol2.0~\cite{NNPDFpol20} and taking the standard deviation, $\delta A_{LL}^{\rm PDF}=\sigma_{\rm rep}[A_{LL}]$. The total uncertainty is the quadrature sum, $\delta A_{LL}=\sqrt{(\delta A_{LL}^{\rm exp})^{2}+ (\delta A_{LL}^{\rm PDF})^{2}}$.

\begin{figure*}[!htbp]
\centering
\includegraphics[width=\linewidth]{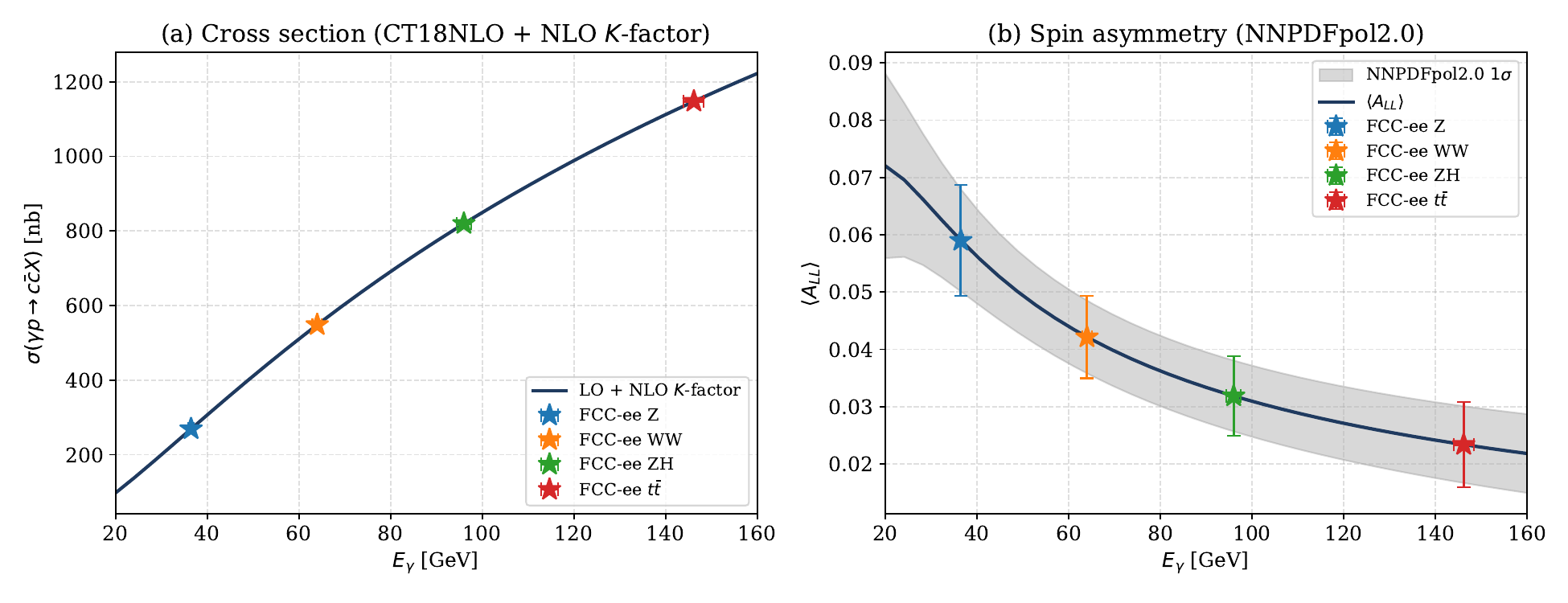}
\caption{(a) Open charm photoproduction cross section
$\sigma(\gamma p\to c\bar{c}X)$ versus $E_{\gamma}$, computed at LO PGF (CT18 NLO) with the NLO $K$-factor $K_\sigma\simeq
1.4$~\cite{BojakStratmann1999}. (b) Longitudinal double-spin asymmetry $\langle A_{LL}\rangle$ using NNPDFpol2.0 with the NLO ratio $K_{\Delta\sigma}/K_\sigma\simeq 0.86$; the shaded band shows the $1\sigma$ NNPDFpol2.0 replica envelope. The four FCC-ee operating points are overlaid with the projected $\delta A_{LL}$ from Table~\ref{tab:asymmetry}.}
\label{fig:asym}
\end{figure*}

The cross sections, asymmetries, event yields and projected uncertainties for each FCC-ee operating mode are summarized in Table~\ref{tab:asymmetry}. For each mode, $\langle A_{LL}\rangle$ and the effective cross section $\sigma$ are obtained as photon-flux-weighted averages over the polarized edge band $E_{\min}<E_{\gamma}<\omega_{\max}$ selected by the pair spectrometer (Sec.~\ref{sec:cbs_simulation}),
\begin{equation}
\langle A_{LL}\rangle =
\frac{\int_{E_{\min}}^{\omega_{\max}} dE_{\gamma}\,n(E_{\gamma})\,
\Delta\sigma(E_{\gamma})}
{\int_{E_{\min}}^{\omega_{\max}} dE_{\gamma}\,n(E_{\gamma})\,
\sigma(E_{\gamma})},
\label{eq:ALL_band}
\end{equation}
where $n(E_{\gamma})=dN_{\gamma}/dE_{\gamma}$ is the simulated CAIN photon spectral density on the edge band, evaluated by histogramming the backscattered photons in 50 energy bins per mode. The corresponding mean photon energy $\langle E_{\gamma}\rangle=(E_{\min}+\omega_{\max})/2$ quoted in the table is used only as a representative value for plotting; the underlying $A_{LL}$, $\sigma$, $\langle x\rangle$ and $\langle\hat{a}_{LL}\rangle$ are all integrated over the full band. Figure~\ref{fig:asym} shows the energy dependence of $\sigma(\gamma p\to c\bar{c}X)$ and $A_{LL}(E_{\gamma})$ across the full FCC-ee photon-energy reach, with the four mode-specific projections superposed.

\begin{table*}[t]
\centering
\caption{Open-charm photoproduction projections per FCC-ee operating mode, integrated over the polarized edge band $E_{\min}<E_{\gamma}<\omega_{\max}$ with the CAIN photon spectrum as weight [Eq.~\eqref{eq:ALL_band}]. The representative photon energy is $\langle E_{\gamma}\rangle=(E_{\min}+\omega_{\max})/2$. Cross sections include the NLO $K$-factor of Ref.~\cite{BojakStratmann1999}, event yields include the geometric acceptance factor $\varepsilon_{\rm geom}=0.85$ for the forward beam hole. The experimental uncertainty $\delta A_{LL}^{\rm exp}$ is the quadrature sum of statistical and systematic contributions; $\delta A_{LL}^{\rm PDF}$ is the NNPDFpol2.0 replica spread; the total uncertainty $\delta A_{LL}$ adds these in quadrature.}
\label{tab:asymmetry}
\begin{ruledtabular}
\begin{tabular}{lccccccccc}
Mode & $\langle E_{\gamma}\rangle$ [GeV] & $\sigma$ [nb] &
$\langle A_{LL}\rangle$ & $N_{\rm event}/$yr &
$\delta A_{LL}^{\rm exp}$ & $\delta A_{LL}^{\rm PDF}$ &
$\delta A_{LL}$ &
$\dfrac{A_{LL}}{\delta A_{LL}^{\rm exp}}$ &
$\dfrac{A_{LL}}{\delta A_{LL}}$ \\
\midrule
$Z$ & 36.5 & 269.4 & 0.0590 & $8.1\times 10^{8}$ & $3.7\times 10^{-3}$ & $8.9\times 10^{-3}$ & $9.7\times 10^{-3}$ & 15.8 & 6.1 \\
$WW$ & 64.0 & 548.5 & 0.0421 & $7.1\times 10^{9}$ & $3.4\times 10^{-3}$ & $6.4\times 10^{-3}$ & $7.2\times 10^{-3}$ & 12.4 & 5.9 \\
$ZH$ & 96.0 & 819.7 & 0.0319 & $1.2\times 10^{9}$ & $3.2\times 10^{-3}$ & $6.2\times 10^{-3}$ & $7.0\times 10^{-3}$ & 9.9 & 4.6 \\
$t\bar{t}$ & 146.2 & 1147.4 & 0.0234 & $3.1\times 10^{8}$ & $3.2\times 10^{-3}$ & $6.7\times 10^{-3}$ & $7.4\times 10^{-3}$ & 7.4 & 3.2 \\
\end{tabular}
\end{ruledtabular}
\end{table*}

The annual integrated charm yield ranges from $3.1\times 10^{8}$ events in the $t\bar{t}$ mode to $7.1\times 10^{9}$ events in the $WW$ mode, exceeding by three to five orders of magnitude the $\sim 10^{5}$ open-charm events accumulated by COMPASS over its full polarized program~\cite{COMPASS2013}. As a direct consequence the statistical uncertainty $\delta A_{LL}^{\rm stat}\sim 0.9$--$4\times 10^{-4}$ remains small compared to the systematic floor in all modes, and the experimental sensitivity is systematics-limited at the few-per-mil level: the experimental significance $A_{LL}/\delta A_{LL}^{\rm exp}$ ranges from $7.4$ in the $t\bar{t}$ mode to $15.8$ in the $Z$ mode. Including the PDF uncertainty from NNPDFpol2.0, the total significance $A_{LL}/\delta A_{LL}$ remains in the $3$--$6\sigma$ range across all four operating modes, with the highest sensitivity in the $Z$ mode.

The measured asymmetry $\langle A_{LL}\rangle$ on the polarized edge band can now be translated into a projection for the polarized gluon distribution and compared against the existing world data on $\Delta g(x)/g(x)$. In the leading-order PGF picture, $\langle A_{LL} \rangle$ is related to the polarized gluon distribution by the standard convolution~\cite{BojakStratmann1999,COMPASS2013}
\begin{equation}
\langle A_{LL}\rangle \;\simeq\;
\langle\hat{a}_{LL}\rangle\,
\frac{\Delta g(\langle x\rangle,\mu^{2})}{g(\langle x\rangle,\mu^{2})},
\label{eq:ALL_to_dgg}
\end{equation}
where $\langle\hat{a}_{LL}\rangle$ is the partonic helicity asymmetry $\Delta\hat{\sigma}/\hat{\sigma}$ averaged over the PGF integrand and the CBS photon spectrum within the band,
\begin{equation}
\langle\hat{a}_{LL}\rangle =
\frac{\int_{E_{\min}}^{\omega_{\max}}\!dE_{\gamma}\,n(E_{\gamma})
      \int_{x_{\min}(E_{\gamma})}^{1}\!dx\,g(x,\mu^{2})\,
      \Delta\hat{\sigma}(xs)}
     {\int_{E_{\min}}^{\omega_{\max}}\!dE_{\gamma}\,n(E_{\gamma})
      \int_{x_{\min}(E_{\gamma})}^{1}\!dx\,g(x,\mu^{2})\,
      \hat{\sigma}(xs)},
\label{eq:aLL_avg}
\end{equation}
and $\langle x\rangle$ is the corresponding integrand-weighted mean gluon momentum fraction, computed by replacing $\hat{\sigma}\to x\hat{\sigma}$ in the numerator of Eq.~\eqref{eq:aLL_avg}. For each FCC-ee operating mode, $\langle\hat{a}_{LL}\rangle$ and $\langle x\rangle$ are evaluated numerically using the CAIN photon histograms of Sec.~\ref{sec:cbs_simulation} and the CT18 NLO gluon distribution; the NLO $K$-factor ratio is folded in as in Eq.~\eqref{eq:ALL_band}. Inverting Eq.~\eqref{eq:ALL_to_dgg} then gives the projected value of the polarized gluon ratio,
\begin{equation}
\frac{\Delta g}{g}(\langle x\rangle)
\;=\;
\frac{\langle A_{LL}\rangle}{\langle\hat{a}_{LL}\rangle},
\qquad
\delta\!\left(\frac{\Delta g}{g}\right)
=
\frac{\delta\langle A_{LL}\rangle}{|\langle\hat{a}_{LL}\rangle|},
\label{eq:dgg_inversion}
\end{equation}
with the experimental and total uncertainties on $\Delta g/g$ obtained by propagating $\delta A_{LL}^{\rm exp}$ and $\delta A_{LL}$ of Table~\ref{tab:asymmetry}. The resulting projections of $\Delta g(\langle x\rangle)/g(\langle x\rangle)$ for each FCC-ee operating mode are listed in Table~\ref{tab:deltag}.

\begin{table*}[t]
\centering
\caption{Projected $\Delta g(\langle x\rangle)/g(\langle x\rangle)$ for each FCC-ee operating mode, derived from $\langle A_{LL}\rangle / \langle\hat{a}_{LL} \rangle$ at the integrand-weighted mean $\langle x\rangle$. The experimental and total uncertainties on $\Delta g/g$ are obtained by propagating $\delta A_{LL}^{\rm exp}$ and $\delta A_{LL}$ of Table~\ref{tab:asymmetry} through Eq.~\eqref{eq:ALL_to_dgg}.}
\label{tab:deltag}
\begin{ruledtabular}
\begin{tabular}{lcccccc}
Mode & $\langle E_{\gamma}\rangle$ [GeV] & $\langle x\rangle$ &
$\langle\hat{a}_{LL}\rangle$ & $\Delta g/g$ &
$\delta(\Delta g/g)_{\rm exp}$ & $\delta(\Delta g/g)_{\rm total}$ \\
\midrule
$Z$ & 36.5 & 0.184 & 0.328 & 0.180 & 0.011 & 0.030 \\
$WW$ & 64.0 & 0.122 & 0.361 & 0.117 & 0.009 & 0.020 \\
$ZH$ & 96.0 & 0.090 & 0.383 & 0.083 & 0.008 & 0.018 \\
$t\bar{t}$ & 146.2 & 0.067 & 0.406 & 0.058 & 0.008 & 0.018 \\
\end{tabular}
\end{ruledtabular}
\end{table*}

\begin{figure*}[!htbp]
\centering
\includegraphics[width=\linewidth]{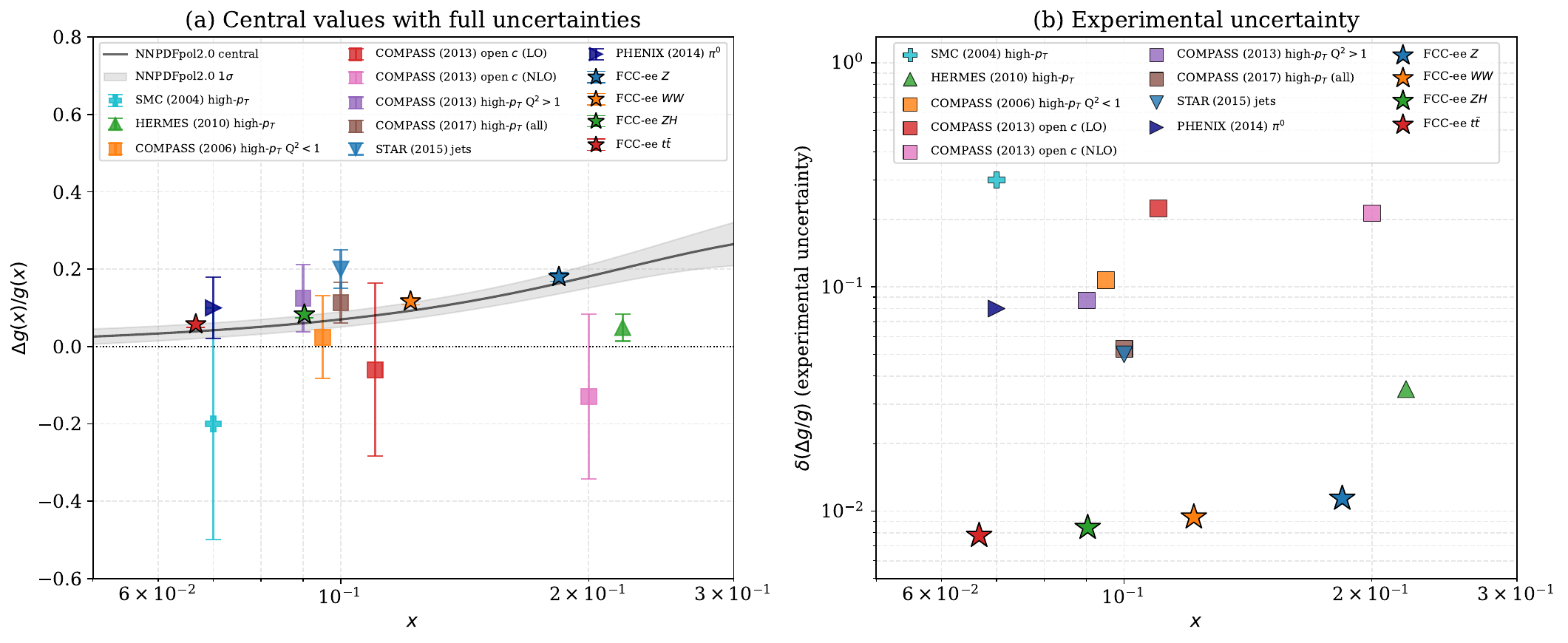}
\caption{(a) Polarized gluon distribution ratio $\Delta g(x)/g(x)$ versus $x$. Markers: world data on direct $\Delta g/g$ extractions (SMC~\cite{SMC2004}, HERMES~\cite{HERMES2010}, five COMPASS results~\cite{COMPASS2013,COMPASS2013NLO,COMPASS2017}) and RHIC pp effective constraints (STAR~\cite{STAR2015}, PHENIX~\cite{PHENIX2014}). The gray band is the $1\sigma$ NNPDFpol2.0 NLO replica envelope~\cite{NNPDFpol20}. Star markers: FCC-ee CBS projections (Table~\ref{tab:deltag}). (b) Experimental uncertainty $\delta(\Delta g/g)$ on a logarithmic scale; the FCC-ee CBS projections lie systematically below the existing world measurements.}
\label{fig:deltag}
\end{figure*}

Figure~\ref{fig:deltag} compares the FCC-ee CBS projections at the four operating points with the world data on direct $\Delta g(x)/g(x)$ measurements. The world dataset consists of seven published results from fixed-target polarized DIS programs: SMC high-$p_{T}$ hadron pairs~\cite{SMC2004}, HERMES high-$p_{T}$ inclusive hadrons~\cite{HERMES2010}, and five COMPASS measurements in the open-charm and high-$p_{T}$ channels (LO and NLO open-charm extractions~\cite{COMPASS2013NLO}; high-$p_{T}$ at $Q^{2}>1$ and $Q^{2}<1$, covering data through 2017~\cite{COMPASS2013,COMPASS2017}). For completeness we also overlay the effective $\Delta g/g$ values extracted from the RHIC pp data, namely STAR inclusive jets at $\sqrt{s}=200$~GeV~\cite{STAR2015} and PHENIX $\pi^{0}$ $A_{LL}$~\cite{PHENIX2014}, at the indicative gluon momentum fractions identified by the DSSV global QCD analysis~\cite{DSSV2014}; these points contribute to the polarized gluon constraint through their inclusion in the NNPDFpol2.0 global fit~\cite{NNPDFpol20} rather than as independent $\Delta g/g$ measurements, and they are reproduced in Fig.~\ref{fig:deltag} for direct visual comparison.

Three features of Fig.~\ref{fig:deltag} are relevant for the assessment of the proposed facility. First, the four FCC-ee CBS projections cover the range $0.07\leq\langle x\rangle\leq 0.19$ with total uncertainties $\delta(\Delta g/g)_{\rm tot}\sim 1.8$--$3.0\times 10^{-2}$, a factor of $\sim 4$--$7$ smaller than the most precise existing direct measurement~\cite{HERMES2010} ($\Delta g/g=0.049\pm0.034({\rm stat}) \pm0.010({\rm sys}\text{-}{\rm exp})^{+0.125}_{-0.099}({\rm sys}\text{-} {\rm models})$, dominated by Monte-Carlo model uncertainties) and a factor of $\sim 7$--$12$ smaller than the COMPASS open-charm result~\cite{COMPASS2013NLO} ($\Delta g/g=-0.13\pm0.15({\rm stat})\pm0.15 ({\rm sys})$ in NLO QCD), which is the only existing extraction in the same partonic process. Second, the present world data display an internal tension at $1$--$2\sigma$: the COMPASS open-charm result prefers a negative central value, while the RHIC pp jet (STAR) and high-$p_{T}$ (COMPASS) results favour a positive central value; the NNPDFpol2.0 global fit, which combines all of these inputs, sits at a small positive central value with the data scatter reflecting the channel-by-channel tensions. No single existing experiment has the statistical or systematic power to resolve this pattern. Third, the FCC-ee CBS projections cluster within the NNPDFpol2.0 band, with uncertainties an order of magnitude below the present spread of the data, so that the proposed measurement would provide an independent, high-statistics test of the open-charm $\Delta g/g$ in the same kinematic regime that currently produces the tension.

The absence of any new direct $\Delta g(x)/g(x)$ measurement program since the completion of the COMPASS analysis in 2017~\cite{COMPASS2017} reinforces this conclusion: the most recent constraints on the polarized gluon enter the global fits exclusively through inclusive $A_{LL}$ measurements at RHIC, with no new event-level $\Delta g/g(x)$ determination on the horizon. The FCC-ee CBS facility proposed here, when realised, would provide four high-statistics direct measurements at four distinct $\langle x\rangle$ in the same kinematic region currently covered by the world data, and would set the dominant new constraint on the polarized gluon distribution in the medium-$x$ region.

\section{Summary and Outlook}
\label{sec:conclusion}
We have presented a complete kinematic and optical design of a polarized gamma-ray facility based on Compton backscattering of intense laser pulses against the FCC-ee electron beams in its full-energy booster, and have evaluated its scientific reach for the measurement of the polarized gluon distribution $\Delta g(x)$ through open-charm photoproduction. The CBS conversion point is located in a dedicated straight section of the FCC-ee full-energy booster, which shares the collider tunnel and operates at all four beam energies, avoiding the complex final-focus optics required at the collider interaction regions. Four laser wavelengths ($199$, $349$, $524$, $797$~nm) saturate the kinematic safety margin $\kappa=4.35$ in the $Z$, $WW$, $ZH$ and $t\bar{t}$ operating modes, yielding photon beams up to $\omega_{\max}=148$~GeV. A flat-beam interaction point saturating $\sigma^{\prime}_{x}=1/\gamma$, with the FSR booster emittances and a laser waist set by the hourglass condition, is adopted throughout, reducing the full-conversion pulse energies to the 45~J--2.9~kJ range. The facility operates on the booster cycle structure, parasitically on the top-up flat-tops ($f_{\rm CBS}=10^{-8}$) or, as baseline, in dedicated extended-flat-top fills in the idle windows ($f_{\rm CBS}=10^{-6}$, beam loss $<1\%$ per cycle); the collider luminosity is unaffected in both, and the operational pulse energies lie in the sub-millijoule to few-millijoule range. CAIN simulations including the full QED treatment of laser--electron interactions are used to characterise the resulting photon spectra, angular distributions and polarization profiles. The interaction-point design holds the electron divergence at $\sigma^{\prime}_{x}=1/\gamma$, so the spectra follow the ideal Klein--Nishina shape for $\kappa=4.35$ and agree with the analytic polarization-transfer function for an unpolarized beam. The downstream collimator system is shown to act as a radiation envelope and shower-cleaning device rather than a polarization filter; the polarized event selection is instead performed event-by-event in the detector by a pair spectrometer that reconstructs $E_{\gamma}$ on the high-energy Compton edge, delivering an effective $P_{\gamma}=0.90$ in the analysed band, the analytic limit for an unpolarized beam. The non-interacting photon flux, with a time-averaged beam power up to $\sim46$~W in the $WW$ mode (dedicated baseline), is absorbed in a downstream beam dump after passing through a forward beam hole in the detector, with a geometric acceptance loss of $\sim 15\%$ included in the event-yield projections.

Convoluting the photon-gluon-fusion cross section, including the NLO $K$-factor of Ref.~\cite{BojakStratmann1999}, with the CT18 NLO and NNPDFpol2.0 gluon distributions and projecting onto a standard NH$_{3}$ DNP polarized target, we obtain annual polarized open-charm event yields between $3.1\times 10^{8}$ in the $t\bar{t}$ mode and $7.1\times 10^{9}$ in the $WW$ mode, exceeding by three to five orders of magnitude the $\sim 10^{5}$ open-charm events accumulated by COMPASS over its full polarized program~\cite{COMPASS2013}. The measurement is systematics-limited in all four operating modes. Consequently, Compton-scattering off the more intense and possibly more focused FCC-ee collider beam instead of the booster beam would not lead to a larger projected precision on $\Delta g/g$, since the measurement is systematics-dominated in all operating modes. The corresponding experimental precision on the longitudinal double-spin asymmetry, $\delta A_{LL}^{\rm exp}\sim 3$--$4\times 10^{-3}$, translates into a projected uncertainty on the polarized gluon distribution of $\delta(\Delta g/g)_{\rm tot}\sim 1.8$--$3.0\times 10^{-2}$ at four distinct values of $\langle x\rangle$ in the medium-$x$ region $0.07\leq\langle x\rangle\leq 0.19$. This precision is a factor of $\sim 3$--$4$ smaller than the experimental uncertainty of the most precise existing direct $\Delta g(x)/g(x)$ measurement (HERMES~\cite{HERMES2010}) and a factor of $\sim 4$--$7$ smaller than its total uncertainty including Monte-Carlo model systematics, and a factor of $\sim 7$--$12$ smaller than the COMPASS open-charm NLO result in the same partonic process~\cite{COMPASS2013NLO}. To our knowledge, no new direct $\Delta g(x)/g(x)$ measurement program is currently on the horizon, and the proposed facility would be the first with the precision required to resolve the channel-by-channel tensions visible in the present world data (Fig.~\ref{fig:deltag}).

The fact that the four FCC-ee operating modes probe distinct but overlapping regions of partonic momentum fraction, with the highest sensitivity reached in the $Z$ mode where both the asymmetry and the photon flux are largest, makes the proposed setup a natural platform for a model-independent extraction of $\Delta g(x)$ in the medium-$x$ region. This range is complementary to the low-$x$ reach of the planned Electron--Ion Collider and extends well beyond the kinematic coverage of present polarized DIS facilities, providing a direct probe of the gluon contribution to the proton spin. Future work will address the inclusion of additional polarized targets (ND$_{3}$, $^{6}$LiD, $^{3}$He) for flavour separation, the use of complementary final states such as $J/\psi$ photoproduction and jet asymmetries, the full NLO treatment of the open-charm cross section beyond the $K$-factor approximation used here, a detailed Geant4 simulation of the detector acceptance and the beam dump, and the optimisation of the laser polarimetry to push $\delta P_{\gamma}/P_{\gamma}$ below the $1\%$ level assumed here. The combination of these refinements with the FCC-ee booster photon flux and parasitic-mode design presented in this work would establish the highest-statistics polarized-photoproduction program ever proposed, and would constitute a decisive contribution to the resolution of the proton spin puzzle.

\section*{Acknowledgments}
The authors would like to thank Umit Kaya and Burak Dagli for useful discussions. AI-assisted tools were used for manuscript preparation and data analysis scripting.

\bibliography{references}

\end{document}